\documentclass[sigconf,screen]{acmart}

\AtBeginDocument{%
  }

\copyrightyear{2026}
\acmYear{2026}
\setcopyright{cc}
\setcctype{by-nc-nd}
\acmConference[CCS '26]{Proceedings of the 2026 ACM SIGSAC Conference on Computer and Communications Security}{November 15--19, 2026}{The Hague, Netherlands}
\acmBooktitle{Proceedings of the 2026 ACM SIGSAC Conference on Computer and Communications Security (CCS '26), November 15--19, 2026, The Hague, Netherlands}
\acmDOI{10.1145/3830454.3846522}
\acmISBN{979-8-4007-2871-6/2026/11}

\usepackage{xspace}
\usepackage[dvipsnames]{xcolor}
\usepackage{tikz}
\usepackage{amsfonts}
\usepackage{amsmath}
\usepackage{subcaption}
\usepackage{tabularray}
\SetTblrStyle{remark}{font=\footnotesize}

\UseTblrLibrary{diagbox}
\usepackage{graphicx}
\usepackage{xurl}
\usepackage{hyperref}
\usepackage[nameinlink]{cleveref}
\usepackage{enumitem}
\usepackage[export]{adjustbox}

\definecolor{LinkBlue}{HTML}{1E88E5}   
\definecolor{DeepBlue}{HTML}{1565C0}   
\definecolor{SeriousBlue}{HTML}{0D47A1}
\definecolor{CalmBlue}{HTML}{1A73E8}   
\definecolor{Navyish}{HTML}{0B3D91}    
\definecolor{MutedBlue}{HTML}{2A6F97}  
\definecolor{Link}{HTML}{2A4B8D} 
\definecolor{Cite}{HTML}{2F7D32} 
\definecolor{URL}{HTML}{3B6EA5}  
\definecolor{BlushRed}{HTML}{F28B82}
\definecolor{NiceGreen}{HTML}{1A7F37}  
\definecolor{NiceRed}{HTML}{B42318}    
\definecolor{TintGreen}{HTML}{EAF4EC}  
\definecolor{TintRed}{HTML}{FBECEA}    
\definecolor{OursRow}{HTML}{E4EBF7}    
\definecolor{TableMuted}{HTML}{9AA0A6} 

\hypersetup{
    pdfborder={0 0 0},
    colorlinks=true,    
    linkcolor=LinkBlue,     
    citecolor=Cite,    
    urlcolor=URL        
}

\crefname{equation}{Eq.}{Eqs.}
\Crefname{equation}{Eq.}{Eqs.}
\crefname{figure}{Fig.}{Figs.}
\Crefname{figure}{Fig.}{Figs.}
\crefname{table}{Tab.}{Tabs.}
\Crefname{table}{Tab.}{Tabs.}
\crefname{section}{\S}{\S\S}
\Crefname{section}{\S}{\S\S}
\creflabelformat{equation}{#2(#1)#3}   
\creflabelformat{figure}{#2#1#3}        
\creflabelformat{table}{#2#1#3}         
\creflabelformat{section}{#2#1#3}       

\crefformat{section}{#2\S#1#3}
\Crefformat{section}{#2\S#1#3}

\crefrangeformat{section}{#3\S#1#4--#5\S#2#6}
\Crefrangeformat{section}{#3\S#1#4--#5\S#2#6}

\usepackage{pifont}
\newcommand{\cmarkgreen}{\textcolor{NiceGreen}{\ding{51}}}
\newcommand{\xmarkred}{\textcolor{NiceRed}{\ding{55}}}

\newcommand{\tgood}[1]{\cmarkgreen\hspace{0.25em}\textbf{#1}}
\newcommand{\tbad}[1]{\xmarkred\hspace{0.25em}#1}
\newcommand{\tna}{\textcolor{TableMuted}{N/A}}
\newcommand{\tsub}[1]{\textcolor{TableMuted}{#1}}

\newcommand{\formulafont}{\small}

\newcommand{\toolname}{\textsc{AimTrap}\xspace}

\newcommand{\CQ}[1]{\ifdefined\SHOWCOMMENTS\textcolor{purple!90}{CQ: #1}\else{}\fi}
\newcommand{\AD}[1]{\ifdefined\SHOWCOMMENTS\textcolor{teal!90}{AD: #1}\else{}\fi}

\newcommand{\jn}[1]{\ifdefined\SHOWCOMMENTS\textcolor{orange!90}{JN: #1}\else{}\fi}

\usepackage{xstring}
\newcommand{\PP}[1]{
\vspace*{1.2pt}
\noindent{\bf \IfEndWith{#1}{.}{#1}{#1.}}
}

\usepackage{tcolorbox}
\newtcolorbox{conclusionbox}{colback=gray!8,colframe=black,width=\linewidth,arc=0.6mm, boxrule=0.4pt, left=1mm,right=1mm,top=1mm,bottom=1mm}

\newcommand*\ttvar[1]{\texttt{\expandafter\dottvar\detokenize{#1}\relax}}
\newcommand*\dottvar[1]{\ifx\relax#1\else
  \expandafter\ifx\string_#1\string_\allowbreak\else#1\fi
  \expandafter\dottvar\fi}

\usepackage{enumitem}

\newcommand*\circled[1]{\tikz[baseline=(char.base)]{
            \node[shape=circle,fill,inner sep=0.8pt] (char) {\textcolor{white}{#1}};}}

\newcommand{\textbi}[1]{\textbf{\textit{#1}}}

\newif\ifshowrevised
\showrevisedtrue
\newcommand{\RevisedOn}{\showrevisedtrue}

\usepackage[normalem]{ulem}

\RevisedOn

\NewTblrTheme{no-caption}{
    \SetTblrTemplate{caption}{empty}
}
\NewTblrTheme{no-caption-head-foot}{
    \SetTblrTemplate{head}{empty}
    \SetTblrTemplate{foot}{empty}
    \SetTblrTemplate{caption}{empty}
}

\tcbuselibrary{breakable, skins}
\newtcolorbox{promptbox}{
    breakable,
    enhanced,
    colback=blue!3,
    colframe=LinkBlue,
    arc=4pt,
    boxrule=1pt,
    left=5pt,
    right=5pt,
    top=5pt,
    bottom=5pt,
    fontupper=\ttfamily\small,
    before upper={\textbf{Prompt: }},
}

\begin{document}

\title{Shoot the Honey, Cloak the Player:\\Towards Zero-Runtime-Overhead Proactive Defense and Detection for Visual Game Cheating}

\author{Jianing Wang}
\orcid{0009-0004-9917-1527}
\affiliation{  \department{School of Cyber Science and Technology}
  \institution{Shandong University}
  \city{Qingdao}
  \country{China}
}
\email{jianingwang@mail.sdu.edu.cn}

\author{Chuqi Zhang}
\orcid{0009-0006-3550-696X}
\affiliation{  \institution{Independent Researcher}
  \city{Hangzhou}
  \country{China}
}
\email{zhangchuqi1999@gmail.com}

\author{Yuancheng Jiang}
\orcid{0009-0006-7833-5208}
\affiliation{  \department{School of Computing}
  \institution{National University of Singapore}
  \city{Singapore}
  \country{Singapore}
}
\email{yuancheng@comp.nus.edu.sg}

\author{Adil Ahmad}
\orcid{0009-0002-4097-3205}
\affiliation{  \department{School of Computing and Augmented Intelligence}
  \institution{Arizona State University}
  \city{Tempe}
  \country{United States}
}
\email{adil.ahmad@asu.edu}

\author{Shanqing Guo}
\orcid{0000-0003-3367-0951}
\authornote{Corresponding author.}
\affiliation{
\department{School of Cyber Science and Technology, Shandong Key Laboratory of Artificial Intelligence Security, State Key Laboratory of Cryptography and Digital Economy Security}
\institution{Shandong University}
  \city{Qingdao}
  \country{China}
}
\email{guoshanqing@sdu.edu.cn}

\begin{abstract}

Visual aimbots have emerged as a serious cheating threat in first-person shooter (FPS) games, as they evade existing anti-cheat defenses by operating only on rendered frames rather than game memory.
However, existing defenses fail to provide an end-to-end solution: post-hoc behavior detectors cannot protect match integrity in real time and are increasingly fragile against human-mimicking aimbots, while proactive runtime defenses often lack accountability, incur substantial overhead, or require intrusive system integration.

We present \toolname, the first end-to-end visual-aimbot defense that combines runtime protection with post-game detection through two adversarial texture mechanisms.
Adversarial Camouflage Textures (ACT) hide real players from aimbots, while Adversarial Honeypot Textures (AHT) lure aimbots into locking onto fake targets, yielding strong evidence of cheating.
\toolname integrates differentiable rendering with Expectation over Renderings for robust 3D texture synthesis and analyzes honeypot-interaction trajectory to facilitate cheating attribution.
In real-game evaluation against a visual aimbot, ACT achieves 85.1\% defense success, AHT achieves 96.9\%.
Compared with prior baselines, \toolname also achieves extremely low false-positive rates with negligible runtime overhead, demonstrating a practical end-to-end defense.
\end{abstract}



\maketitle

\section{Introduction}
\label{s:introduction}

Global video game revenue is projected to reach \$564 billion by 2026~\cite{video-game-market}, with first-person shooters (FPS) remaining one of the most popular and lucrative markets~\cite{video-game-percentage}.
Unfortunately, FPS games are persistently plagued by cheating, which undermines fair competition, degrades player experience, reduces retention, and causes substantial financial losses for the gaming industry~\cite{cheating-impact,cheating-impact2,Games24-unpacking,pathways-into-cybercrime}. 
Among these threats, \emph{aimbots} are particularly damaging in competitive FPS games because they automatically aim and shoot opponents to provide cheaters with an unfair advantage~\cite{cloak,hawk,xguardian}.

Aimbots broadly fall into two categories: memory-based and visual-based.
\emph{Memory-based aimbots} directly read or manipulate game memory to recover precise enemy locations.
They are increasingly constrained by existing anti-cheat systems based on memory monitoring, kernel-level protection, and trusted execution support~\cite{vac,eac,battleye,vanguard,botscreen,blackmirror}.
In contrast, \emph{visual-based aimbots}~\cite{rootkit_aimbot,sunoner_aimbot,aimr,apex_aimbot} operate solely on rendered display frames and therefore evade these defenses by never tampering with game memory.
Powered by advanced object detection models such as YOLO~\cite{yolo,faster-rcnn,nanodet}, they can achieve superhuman aiming precision and speed while remaining highly stealthy.

Existing defenses against visual-based aimbots either perform \emph{post-hoc behavior detection} or implement \emph{proactive runtime defense}.
Post-hoc detectors~\cite{CCNC06-Bayesian, CIG11-ML-Detection, CIG13-Behavioral, PALO15-Bot-Detection, DSN17-AimDetect, ICONIP12-Comparison-Matrix, IJCNN12-Statistical, SIGMETRICS16-SVM-Aimbot, ML21-Multivariate, CIIS20-AITools, TS24-vadnet, hawk,xguardian} analyze gameplay statistics to identify suspicious behavior after a match, whereas runtime defenses~\cite{cloak,v-IPCCC23-Perturbation,advmap} intervene during gameplay by perturbing rendered visuals to mislead the aimbot.
However, they face the following fundamental limitations:

\noindent\textbf{L1. Lack of both accountability and prevention.}
Neither approach delivers an end-to-end solution.
Post-hoc detectors offer \emph{accountability without prevention}: they identify cheating only after unfair advantage has already been gained and therefore cannot protect match integrity in real time.
Runtime defenses provide \emph{prevention without accountability}: although they may disrupt cheating during play, they cannot provide forensic evidence for post-game enforcement, lack the ability to confidently attribute misconduct, and thus fail to penalize cheaters effectively.

\AD{L2: Fragility against human-mimicking aimbots.} \jn{solved}
\noindent\textbf{L2. Fragile against advanced human-mimicking aimbots.}
Visual aimbots do not exhibit many explicit unnatural behaviors of memory-based cheats, such as pre-firing or perfect tracking enemies through obstacles, which existing detectors rely on, as they share the same field of vision as humans.
Moreover, modern cheats employ advanced strategies to mimic human aiming behavior, such as mouse smoothing~\cite{rootkit_aimbot, sunoner_aimbot, aimr}, GAN-based trajectory generation~\cite{gan-aimbots}, or neuromuscular aim assist~\cite{neuromuscular-aim}.
This further narrows the statistical gap between skilled legitimate players and cheaters.

\AD{Consider merging L3/L4 into "L3: Require expensive system modifications" They are both sub-problems of the same higher-level problem.}
\jn{solved.}
\CQ{I think we may not want to mention trusted environment / SGX here. Our model also assumes a trusted host OS.}
\jn{solved.}

\noindent\textbf{L3. Face practical deployment barriers in real FPS games.}
Prior defenses face three practical barriers.
First, display-perturbation defenses impose substantial per-frame overhead, making them unsuitable for latency-sensitive gameplay~\cite{cloak,v-IPCCC23-Perturbation}.
Second, texture-perturbation defenses~\cite{advmap} offer limited geometric coverage: they can only apply to 2D planar surfaces and cannot protect general geometry, such as 3D player and non-planar building models.
Third, many defenses require intrusive integration into the game stack, including high-frequency game data collection, peripheral logging, or hooks into shaders and graphics APIs~\cite{CIG13-Behavioral,ML21-Multivariate,blackmirror,botscreen,cloak,v-IPCCC23-Perturbation}.
Together, these barriers limit their portability, maintainability, and adoption in production games.

These limitations raise a key question in the defense landscape:
{can a visual anti-cheat solution simultaneously prevent cheating online, produce actionable post-game evidence, and remain lightweight enough for deployment in modern FPS games?}

We answer this question affirmatively with two key insights.
First, \textbf{we transform visual-aimbot detection from passive behavior classification into an active visual challenge-response problem}.
Rather than inferring cheating from noisy and increasingly human-like aiming behavior, we introduce {external visual signals that are salient to visual aimbots but unlikely to affect benign players}.
Second, unlike display-space perturbations that require expensive per-frame processing or prior texture-space defenses that are limited to 2D planar surfaces, \textbf{we shift the cost to an offline, 3D-aware adversarial texture synthesis stage}.

\begin{figure}[t!]
    \centering
    \includegraphics[width=1.0\linewidth]{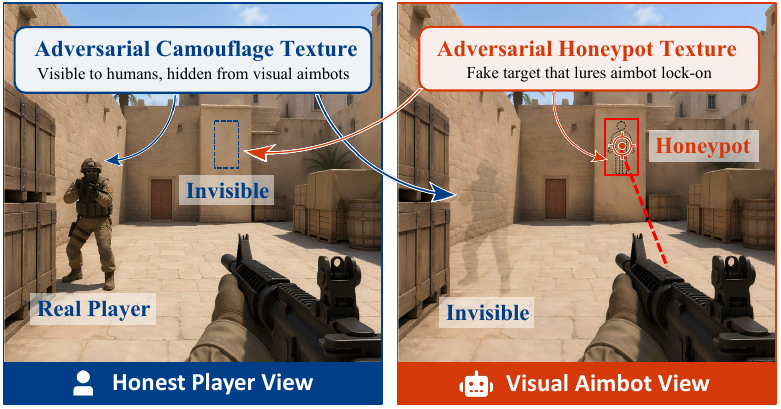}
    \caption{A high-level illustration of \toolname's principles.}
    \label{f:illustration}
\end{figure}

To instantiate these insights, we propose two complementary mechanisms~(\cref{f:illustration}).
\emph{Adversarial Honeypot Textures (AHT)} are static decoys placed in game scenes.
They appear as high-confidence, player-like objects \emph{only} to visual aimbots, inducing deliberate aiming behaviors (i.e., lock-on), while remaining non-salient to honest players.
\emph{Adversarial Camouflage Textures (ACT)} are static noises applied to player models that suppress the aimbot's detection confidence while remaining visually natural to humans.
Together, ACT reduces aimbot's ability to detect real opponents, while AHT redirects it toward non-existent targets; \emph{repeated lock-on to such targets then becomes strong evidence of visual cheating}.

Realizing an end-to-end solution based on proposed mechanisms poses two technical challenges.
\textbf{C1.} It is non-trivial to synthesize \emph{robust} adversarial textures (AHT/ACT) that survive in the 3D rendering pipeline, including UV mapping~\cite{uv_mapping}, viewpoint changes, and lighting variations.
In addition, secure and scalable texture management is also challenging, as textures must be protected from client-side adversaries and deployed efficiently across millions of online matches.
\textbf{C2.} Even with honeypot textures, reliably detecting cheating with high accuracy and low false-positive rates remains difficult.
The key challenge is to precisely identify sparse honeypot interactions from massive gameplay events while remaining robust to noisy interactions from honest players.

To this end, we present \toolname, a framework that both \emph{proactively defends against} visual aimbots at runtime and \emph{detects} their presence with \emph{strong and explainable evidence}, while preserving practical deployability in real-world FPS games.
To enable robust texture synthesis (\textbf{C1}), \toolname integrates an end-to-end differentiable rendering pipeline and adopts Expectation over Renderings (EoR) to optimize texture-space perturbations under diverse rendering conditions (\cref{s:act}, \cref{s:aht}).
It further incorporates strategies for secure texture management (\cref{s:texture_management}) and efficient deployment (\cref{s:texture_deployment}).
To enable effective and reliable cheating detection (\textbf{C2}), \toolname designs a honeypot-interaction mouse trajectory extraction and abnormal honeypot-interaction identification algorithm (\cref{s:detection}).

We evaluate \toolname against a state-of-the-art visual aimbot~\cite{rootkit_aimbot} in real-game settings with real textures, replay logs, and environments.
\toolname achieves strong proactive defense, with ACT reaching an average success rate of 85.1\% and AHT 96.9\% across diverse rendering conditions (\cref{e:defense_effectiveness}, \cref{e:robustness}).
AHT also transfers to unseen aimbot architectures, reaching 86.9\% average decoy success rate across seven detectors when optimized with an ensemble proxy and transfer-enhancing techniques (\cref{e:transferability}).
Meanwhile, its cheating detector identifies abnormal honeypot interactions with 98.05\% accuracy and 98.98\% recall, and correctly classifies all 40 real-game matches without falsely flagging any of the 20 benign matches (\cref{e:trajectory_identification}, \cref{e:real-world_detection}).
By aggregating honeypot-interaction evidence at the player level, \toolname further suppresses false positives: requiring more abnormal interactions and consistent evidence across multiple matches reduces the estimated false-positive rate by orders of magnitude (\cref{e:trajectory_identification}).
The detector also outperforms the state-of-the-art \textsc{XGuardian}~\cite{xguardian}, which complements \toolname by providing broader coverage of aimbots and wallhacks.
Compared with prior perturbation-based defenses~\cite{cloak,v-IPCCC23-Perturbation}, \toolname incurs negligible runtime overhead, while preserving perceptual quality and practical deployability (\cref{e:defense_effectiveness}, \cref{e:gaming_impact}, \cref{e:overhead}).
These results demonstrate that \toolname is a practical end-to-end defense against visual aimbots.

\noindent\textbf{Contributions.}
This paper makes the following contributions:

(1) We introduce a novel and reliable signal for distinguishing visual aimbots from benign players: adversarial honeypots that induce cheat-specific lock-on behavior.

(2) We present \toolname, the first end-to-end defense against visual aimbots that unifies proactive runtime protection and post-game detection through two complementary and effective adversarial texture mechanisms, without introducing runtime overhead.

(3) We evaluate \toolname in real-game environments against an open-source visual aimbot, demonstrating superior effectiveness, explainable detection signals, and practical deployability compared to state-of-the-art baselines.

\section{Background and Motivation}
\label{s:motivation}

\subsection{Visual Aimbot: A New Threat in FPS Games}
\label{s:aimbot}

Unlike traditional memory-based aimbots, visual aimbots~\cite{rootkit_aimbot,sunoner_aimbot,aimr,apex_aimbot} achieve superhuman precision without tampering with game memory.
Instead, they capture real-time display frames and process them through high-speed object detection models, such as YOLO, Faster R-CNN, and NanoDet, to identify on-screen enemies and automatically lock onto targets~\cite{yolo,faster-rcnn,nanodet}.
This vision-based paradigm offers three distinct advantages over memory-based aimbots.

\textit{Stealthiness via system-level evasion.}
Visual aimbots can bypass conventional memory-based anti-cheats~\cite{vac,eac,battleye,vanguard} as they operate only on legitimate video output.
They can also avoid kernel-level monitoring~\cite{battleye,vanguard} through two-PC setups, where an external machine processes frames through a capture card and returns control signals via simulated peripherals~\cite{swisschili}.
Recent proof-of-concept systems further show that visual targeting can be coupled with Electrical Muscle Stimulation (EMS) to induce rapid user input, making cheating behavior appear closer to organic human control~\cite{neuromuscular-aim}.

\textit{Adaptability}.
Their vision pipeline is inherently game-agnostic.
They can seamlessly transfer across different games, platforms, and even cloud gaming environments~\cite{xbox-cloud-gaming,geforce-now,playstation}, where traditional memory-based aimbots are infeasible.

\textit{Low barrier to entry}.
Visual aimbots eliminate the need for complex, continuous reverse engineering of obfuscated game binaries.
Open-source vision models and LLM-assisted development tools further lower the technical barrier, enabling individuals with minimal technical expertise to develop and deploy effective cheats.

\subsection{Existing Defenses Against Visual Aimbots}
\label{s:existing-work}

Existing defenses against visual aimbots can be categorized into the following two paradigms, summarized in \cref{t:existing_work}.

\noindent
\textbf{Post-hoc behavior detection.}
Such detectors~\cite{CCNC06-Bayesian, CIG11-ML-Detection, CIG13-Behavioral, PALO15-Bot-Detection, DSN17-AimDetect, ICONIP12-Comparison-Matrix, IJCNN12-Statistical, SIGMETRICS16-SVM-Aimbot, ML21-Multivariate, CIIS20-AITools, TS24-vadnet, hawk, xguardian} analyze gameplay statistics to detect cheating behavior after a match.
The statistics can be collected from in-game or system-level data, or extracted from post-game replay logs.
They analyze these statistics using machine learning techniques to identify suspicious behavior.
For instance, \textsc{Hawk}~\cite{hawk} requires game-specific features (e.g., flash bang affection duration on opponents and teammates) to evaluate player behavior, while \textsc{XGuardian}~\cite{xguardian} relies on pitch and yaw angle data to inspect aiming trajectories.
Additionally, \textsc{XGuardian} explicitly introduces explainability into its predictions via SHAP~\cite{xguardian}, whereas other approaches struggle to provide clear evidence for their decisions.

\noindent
\textbf{Proactive runtime defenses.}
Such approaches~\cite{cloak,v-IPCCC23-Perturbation,advmap} directly interfere with the aimbot's vision functionality during gameplay.
In particular, display-perturbation methods~\cite{cloak,v-IPCCC23-Perturbation} inject imperceptible adversarial noise into each rendered frame to suppress the functionality of aimbots, but incur significant performance overhead.
For instance, Invisibility Cloak~\cite{cloak} applies camouflage perturbations to hide real players. 
Similarly, Nhu et al.~\cite{v-IPCCC23-Perturbation} combine camouflage perturbations with decoy perturbations (injecting fake players) to degrade the cheat's accuracy.
Alternatively, texture-perturbation approaches~\cite{advmap} apply noises directly within the texture space to avoid per-frame computational costs.

\begin{table*}[t]
\centering
\begin{talltblr}[
    caption = {Comparison between \toolname (this work) and state-of-the-art visual anti-cheat solutions.\AD{It would not be easy for reviewers to map between this table and L1-L3. Make it clearer.}\jn{solved.}}, 
    remark{Note} = {\emph{Evidence}: whether the method provides human-interpretable evidence for worensic analysis or manual inspection. \emph{Anti-Mimic}: denotes resilience against advanced aimbots designed to mimic human aiming behavior~\cite{gan-aimbots,neuromuscular-aim}. \emph{Deployment}: whether the method can be deployed transparently in practice without requiring intrusive changes.},
    label = {t:existing_work}
]{
    stretch = 0.60,
    colspec = {l|c|ccc|cccc|c},
    row{1} = {font=\scriptsize\bfseries},
    row{2} = {font=\scriptsize\bfseries},
    row{3-Z} = {font=\scriptsize},
    row{9} = {bg=OursRow, fg=black},
}
\hline[1pt]
\SetCell[r=2]{l} System &
\SetCell[r=2]{c} Category &
\SetCell[c=3]{c} Post-hoc Behavior Detection & & &
\SetCell[c=4]{c} Proactive Runtime Protection & & & &
\SetCell[r=2]{c} Deployment \\
\hline

& & Input & Evidence & Anti-Mimic
& Space & Camouflage & Decoy & Overhead & \\
\hline

BotScreen~\cite{botscreen} (SEC'23)
& Detection
& \tbad{Gamedata}
& \xmarkred
& \xmarkred
& \tna
& \tna
& \tna
& \tna
& \xmarkred\,\tsub{(Legacy Intel SGX)} \\

Nhu et al.~\cite{v-IPCCC23-Perturbation} (IPCCC'23)
& Protection
& \tna
& \tna
& \tna
& \tbad{Display output}
& \cmarkgreen
& \cmarkgreen
& \tbad{Severe}
& \xmarkred\,\tsub{(Display modification)} \\

Invisibility Cloak~\cite{cloak} (SEC'24)
& Protection
& \tna
& \tna
& \tna
& \tbad{Display output}
& \cmarkgreen
& \xmarkred
& \tbad{High}
& \xmarkred\,\tsub{(Display modification)} \\

AdvMap~\cite{advmap} (TG'25)
& Protection
& \tna
& \tna
& \tna
& \tbad{2D texture}
& \xmarkred
& \cmarkgreen
& \tgood{Negligible}
& \cmarkgreen \\

\textsc{Hawk}~\cite{hawk} (TIFS'25)
& Detection
& \tgood{Replay}
& \xmarkred
& \xmarkred
& \tna
& \tna
& \tna
& \tna
& \cmarkgreen \\

\textsc{XGuardian}~\cite{xguardian} (SEC'26)
& Detection
& \tgood{Replay}
& \cmarkgreen
& \xmarkred
& \tna
& \tna
& \tna
& \tna
& \cmarkgreen \\

\textbf{\toolname} (this work)
& \textbf{Both}
& \tgood{Replay}
& \cmarkgreen
& \cmarkgreen
& \tgood{3D texture}
& \cmarkgreen
& \cmarkgreen
& \tgood{Negligible}
& \cmarkgreen \\

\hline[1pt]
\end{talltblr}
\end{table*}

\subsection{Limitations of Existing Defenses} 
\label{s:limitations}

We compare \toolname with state-of-the-art anti-cheat solutions in \cref{t:existing_work}, identifying critical limitations motivating our approach:

\noindent\textbf{L1. Lack of both accountability and prevention.}
Neither post-hoc behavior detection nor proactive runtime protection delivers an end-to-end anti-visual-cheating solution (\cref{t:existing_work}).
Post-hoc detectors offer \emph{accountability without prevention}: they identify cheating only after an unfair advantage has been gained, failing to protect match integrity in real time.
Runtime defenses provide \emph{prevention without accountability}: although they may disrupt cheating during play, they typically cannot provide reliable forensic evidence for post-game enforcement, lack the ability to confidently attribute misconduct, and thus fail to penalize cheaters effectively.

\noindent\textbf{L2. Existing detectors are fragile against advanced visual aimbots.}
Visual aimbots can easily evade conventional signature-based detection~\cite{vac,eac,battleye,vanguard} (\cref{s:aimbot}).
They also challenge behavioral detectors~\cite{CIG13-Behavioral, CIG11-ML-Detection, PALO15-Bot-Detection, DSN17-AimDetect, ICONIP12-Comparison-Matrix, CCNC06-Bayesian, IJCNN12-Statistical, SIGMETRICS16-SVM-Aimbot,hawk,xguardian} for two primary reasons.
First, they do not exhibit explicit unnatural behavior patterns of memory-based cheats, such as pre-firing or perfectly tracking enemies through obstacles.
Second, they employ advanced strategies to mimic human aiming trajectories, such as mouse smoothing~\cite{rootkit_aimbot, sunoner_aimbot, aimr} and GAN-based generation~\cite{gan-aimbots}, making them difficult to distinguish from highly-skilled legitimate gameplay. 
Moreover, neuromuscular aim assist~\cite{neuromuscular-aim} (\cref{s:aimbot}) produces aiming patterns physiologically consistent with natural muscle responses, further blurring the line between human and machine behavior.

\AD{I'd move the explanation from 3.1 to here. This is too short and looks weird. I think you can remove section 3.1 and merge it fully here.}
\AD{Consider merging and shortening L3/L4 as I suggested in the intro.}
\jn{solved.}

\begin{table}[t]
\centering
\begin{talltblr}[
    caption = {Runtime overhead comparison of \toolname and prior proactive runtime protection solutions.},
    remark{Note} = {N/R denotes hardware not reported. FPS values for \cite{v-IPCCC23-Perturbation, cloak} come from their papers; \cite{advmap} and \toolname use online benchmarks, while CPU/memory may differ.},
    label = {t:overhead_comparison}
]{
    width = \linewidth,
    stretch = 0.60,
    colspec = {X[2.25cm,l] X[1.85cm,l] X[1.15cm,c] X[1.55cm,c] X[1.05cm,r]},
    colsep = 1.5pt,
    row{1} = {font=\scriptsize\bfseries},
    row{2-Z} = {font=\scriptsize},
    hline{1,Z} = {1pt},
    hline{2} = {0.5pt}
}
    System / Baseline & Defense Space & GPU & Resolution & FPS \\
    \SetCell[c=2]{l} Gameplay baseline (Casual -- Esports) & & -- & $1920 \times 1080$ & $144\text{--}360$ \\
    \SetCell[c=2]{l} Base hardware baseline (Steam survey~\cite{steam-hwsurvey}) & & RTX 3060 & $1920 \times 1080$ & $150\text{--}350$ \\
    \SetCell[c=2]{l} High-end GPU hardware baseline & & RTX 4090 & $1920 \times 1080$ & $400\text{--}600$ \\
    \hline[dashed]
    Nhu et al.~\cite{v-IPCCC23-Perturbation} & Display output & N/R & $1920 \times 1080$ & $\sim 0.12$ \\
    Invisibility Cloak~\cite{cloak} & Display output & RTX 4090 & $320 \times 320$ & $21\text{--}77$ \\
    AdvMap~\cite{advmap} & 2D texture & RTX 4090 & $1920 \times 1080$ & $400\text{--}600$ \\
    \textbf{\toolname} (this work) & 3D texture & RTX 4090 & $1920 \times 1080$ & $400\text{--}600$ \\
\end{talltblr}
\end{table}

\noindent\textbf{L3. Face practical deployment barriers in real FPS games.}
First, display-perturbation defenses~\cite{v-IPCCC23-Perturbation,cloak} struggle to meet the strict performance standards of real-time FPS games ($144\text{--}360\,\mathrm{fps}$)~\cite{fps1,fps2}.
As shown in \cref{t:overhead_comparison}, Nhu et al.~\cite{v-IPCCC23-Perturbation} require 8.5 seconds per frame, while Invisible Cloak~\cite{cloak} achieves only $21\text{--}77\,\mathrm{fps}$ on a high-end RTX 4090 (extrapolating to $9\text{--}34\,\mathrm{fps}$ on commodity RTX 3060s~\cite{steam-hwsurvey,rtx3060-4090}).
Second, texture-perturbation defense~\cite{advmap} reduces runtime overhead but exhibits limited generalizability.
Specifically, AdvMap~\cite{advmap} is restricted to 2D planar surfaces and thus cannot be extended to general 3D geometry, including player models and non-planar building models.
Last, integrating prior approaches requires disruptive engine or system changes.
Behavior-based detectors~\cite{CIG13-Behavioral,ML21-Multivariate,botscreen} depend on high-frequency game data collection, peripheral logging, or specialized hardware support (e.g., the legacy Intel SGX).
Furthermore, display-perturbation approaches~\cite{cloak,v-IPCCC23-Perturbation} require deep integration with the rendering pipeline, such as hooking shaders or graphics APIs.
These intrusions complicate compatibility and maintainability of off-the-shelf games.

\AD{Try to keep each L1-L3 roughly the same size. Looks much better visually and will not make the reviewer think that one limitation is greater than the other.}
\jn{solved.}

\subsection{Problem Statement and Threat Model}
\label{s:problem_statement}

\AD{This can be its own section }

This paper addresses the emerging threat of \emph{visual aimbots} in FPS games with a defense that (i) proactively disrupts them during gameplay and (ii) provides reliable evidence to identify cheaters.

\PP{Assumptions.}
We assume that standard anti-cheat mechanisms (e.g., EAC and Vanguard) already protect the game client against memory tampering and code injection~\cite{vac,eac,vanguard}. 
We trust the integrity of the game engine, operating system, and network stack.
Thus, we specifically focus on external, vision-based cheats.

\noindent\textbf{Adversary Capabilities.}  
Consistent with today's common practice, cheaters have the following capabilities:  
\textit{(1) Frame Capture}: Accessing rendered game frames directly or via external hardware (e.g., HDMI capture cards) at runtime to bypass signature-based detectors.
\textit{(2) Input Injection}: Generating synthetic mouse and keyboard inputs using OS-level simulation or external USB emulators.
\textit{(3) Computer Vision}: Applying models (e.g., YOLO \cite{yolo}) to captured frames for superhuman aiming precision.
\textit{(4) Behavioral Mimicry}: Post-processing aiming trajectories and adding natural delays to evade behavioral detection \cite{gan-aimbots,neuromuscular-aim}.

\AD{Idea: consider making two subsections here titled "Adversarial Honeypot" and "Adversarial Camouflage". Then, explain what and why in each of these sections.}

\section{Approach}
\label{s:approach}

\subsection{Enabling Observations}
\label{s:enabling_observations}

Our approach is motivated by three FPS gameplay observations.

\noindent\textbf{Observation 1.}
\textit{Visual aimbots react to detector confidence rather than human semantic understanding.}
Visual aimbots operate on rendered frames and rely on object detectors to identify targets.
Therefore, they can react to detector-salient visual cues even when those cues do not correspond to meaningful enemies from a human player's perspective.
This creates an opportunity to manipulate aimbots through visual cues that are recognizable to detectors but imperceptible to legitimate players.

\begin{figure}[t!]
    \centering
    \includegraphics[width=0.9\linewidth]{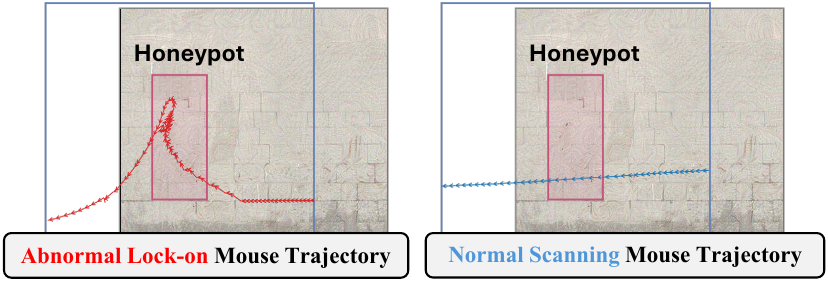}
    \caption{Honeypot-interaction trajectory comparison.}
    \label{f:trajectory}
\end{figure}

\noindent\textbf{Observation 2.}
\textit{If a detector-salient but human-imperceptible visual cue induces repeated lock-on, that behavior becomes a cheat-specific signal.}
We define such cues as \emph{visual honeypots}.
\cref{f:trajectory} illustrates the contrast between normal players and visual aimbots.
Since normal players cannot perceive honeypots as meaningful targets, they do not intentionally aim at those locations.
Their mouse trajectories either pass smoothly across honeypot regions, like scanning, or remain relatively stable at specific points due to normal pre-aiming.
In contrast, aimbots are attracted to honeypots and repeatedly lock onto them, exposing automated target selection.

\begin{figure*}[t]
    \centering
    \includegraphics[width=0.95\linewidth]{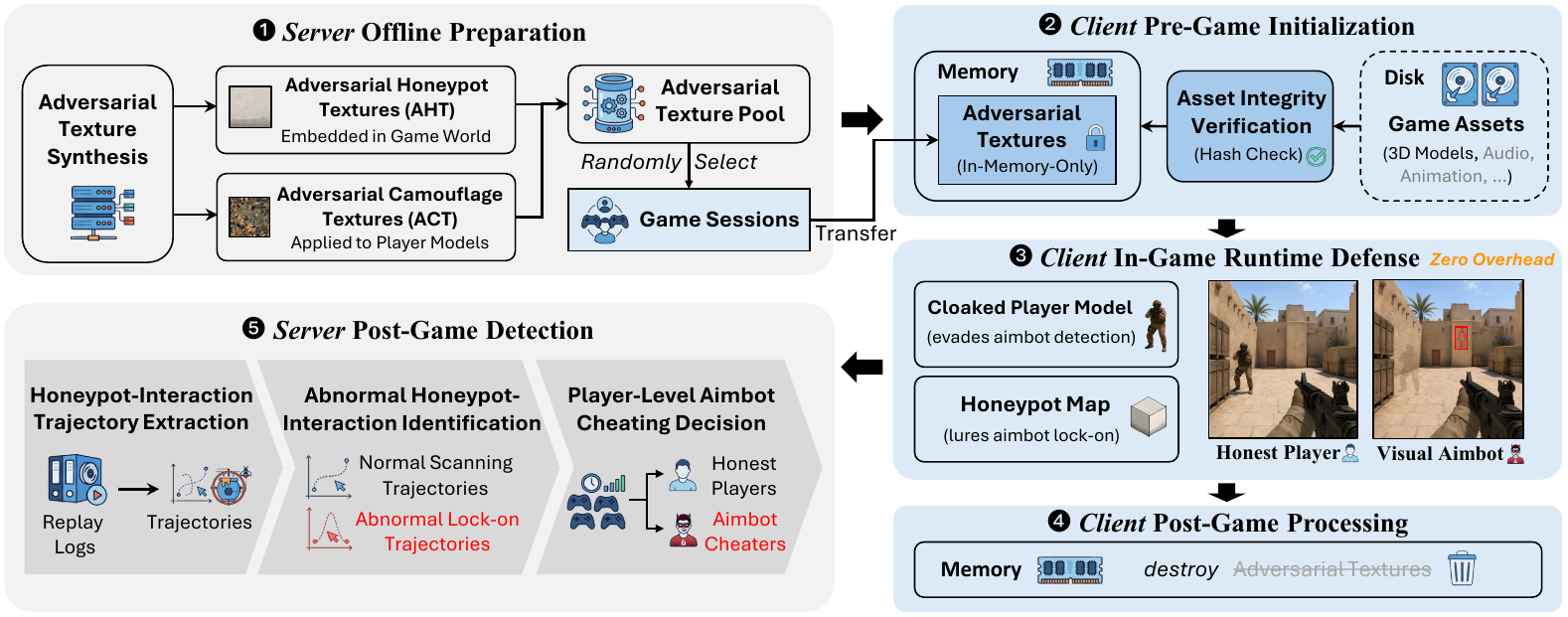}
    \caption{Overview of \toolname.}
    \label{f:overview}
\end{figure*}

\noindent\textbf{Observation 3.}
{\em Relative motion makes static cues dynamic aimbot targets.}
FPS gameplay is inherently dynamic: 
as players move or rotate their view, the perspective projection of the 3D scene continuously changes.
By the principle of relative motion, static honeypots continuously shift in screen space and can repeatedly enter aimbot's detection region around the crosshair.
This makes static honeypots behave like dynamic fake targets from aimbot's perspective, without runtime animation or rendering modification.

\subsection{Adversarial Texture Mechanisms}
\label{s:key_insight_approach}

Building on these observations, we present two key insights.

\noindent \textbf{Insight 1.}
\textit{We transform visual-aimbot detection from passive behavior classification into an active visual challenge-response problem.}
Rather than inferring cheating only from noisy and increasingly human-like aiming behavior, we introduce detector-salient visual signals that are unlikely to affect legitimate players.

\noindent \textbf{Insight 2.}
\textit{Static game assets provide a practical carrier for 3D-aware adversarial visual cues.}
By embedding adversarial cues directly into textures, we can synthesize them in an offline, 3D-aware pipeline and preserve their detector salience across diverse rendering conditions, avoiding expensive per-frame runtime perturbation.

We therefore propose two complementary mechanisms~(\cref{f:illustration}).

\textbf{Adversarial Honeypot Textures (AHT)} contain static decoys, i.e., honeypots, which are placed in game scenes.
They are optimized to lure aimbots into targeting non-existent enemies while remaining visually non-salient to legitimate players.
AHT degrades cheat effectiveness by redirecting aimbot precision toward irrelevant targets and provides explainable post-game evidence through abnormal lock-on behavior.

\textbf{Adversarial Camouflage Textures (ACT)} apply static perturbations to player models.
They suppress the aimbot's detection confidence while remaining visually natural to humans, inspired by prior display-perturbation defenses~\cite{cloak,v-IPCCC23-Perturbation}.
ACT reduces the aimbot's ability to engage real players and complements AHT by redirecting target selection toward visual honeypots.

Together, AHT and ACT degrade visual aimbots at runtime while enabling post-game cheat detection, addressing \textbf{L1}.
Unlike detectors that infer cheating from noisy human-mimic aiming behavior, our visual honeypots introduce an explicit in-game signal: repeated, intentional lock-on to human-imperceptible decoys, which is cheat-specific and unlikely to arise from benign play, addressing \textbf{L2}.
Both AHT and ACT are synthesized offline and rendered as normal assets, requiring neither per-frame computation nor invasive display-pipeline integration.
Additionally, detection is performed server-side using standard in-game replay logs already available in modern FPS games, addressing \textbf{L3}.

\subsection{Challenges}
\label{s:challenges}

Realizing an end-to-end solution presents two technical challenges.

\noindent\textbf{C1: Robust adversarial texture synthesis and management.}
Synthesizing \emph{static} 2D textures that remain adversarial after 3D game rendering is challenging.
In games, textures are mapped onto 3D meshes and processed by complex rendering pipelines, while viewpoints and distances vary unpredictably across players.
Thus, a single perturbation must remain effective under diverse rendering conditions encountered in practice~(\cref{s:act}, \cref{s:aht}).
Moreover, since these textures are deployed on the client, adversaries can mount adaptive attacks, for example by probing to infer honeypot locations or extracting the textures to retrain their aimbots for increased robustness.
This requires a secure and efficient texture management mechanism that limits attacker predictability~(\cref{s:texture_management}, \cref{s:texture_deployment}).

\noindent\textbf{C2: Effective honeypot-based detection.}
Designing a detector that uses honeypot interactions as reliable evidence without causing false positives is also challenging.
The system must locate, aggregate, and interpret sparse interactions within massive replay datasets---often the proverbial needle in a haystack.
In addition, benign players may occasionally trigger such interactions by chance.
Therefore, the detector must tolerate rare benign events while still confidently identifying cheat-driven behavior~(\cref{s:detection}).

\section{System Overview}
\label{s:overview}

\toolname is an anti-visual-cheat framework that proactively interferes with aimbots at runtime with zero overhead and detects aimbot cheating behavior.
To address outlined challenges (\cref{s:challenges}), \toolname consists of five key phases as illustrated in \cref{f:overview}:

\noindent
\circled{1} \textbf{\emph{Server} Offline Preparation}.
The server synthesizes a diverse pool of adversarial textures offline (\cref{s:synthesis}), including
(a) Adversarial Camouflage Textures (\cref{s:act}) for player models to evade aimbot targeting, and
(b) Adversarial Honeypot Textures (\cref{s:aht}) embedded in the game world to lure aimbots and expose their cheating behavior.
We randomize the texture synthesis and per-session selection to limit attacker predictability (\cref{s:texture_management}).

\noindent
\circled{2} \textbf{\emph{Client} Pre-Game Initialization} (\cref{s:texture_management}).
In this phase, the client loads and prepares game assets.
In detail, (a) adversarial textures are securely transmitted from the server to the client's \emph{memory} before the match starts, and
(b) other assets (e.g., 3D models) are loaded from disk and verified against expected hash values.
This integrity verification prevents attackers from tampering with game assets to weaken our adversarial textures.

\noindent
\circled{3} \textbf{\emph{Client} In-Game Runtime Defense.}
At runtime, our adversarial textures proactively disrupt aimbot's functionality: ACT-camouflaged player models evade aimbot detection, while AHT-based honeypots embedded in the map lure aimbot to lock onto them.
As these textures are \emph{3D-robust} and \emph{static} game assets,
\toolname does not need to modify the rendering pipeline at runtime, unlike baselines that inject or alter rendering~\cite{cloak,v-IPCCC23-Perturbation}, and therefore introduces no performance overhead during gameplay (\cref{e:overhead}).

\noindent
\circled{4} \textbf{\emph{Client} Post-Game Processing} (\cref{s:texture_management}).
The client discards all downloaded adversarial textures immediately after the match.
This \emph{in-memory-only} lifecycle places them under existing memory-based anti-cheat protections (see Assumptions in \cref{s:problem_statement}) and prevents any on-disk persistence that could expose them to aimbot developers.

\noindent
\circled{5} \textbf{\emph{Server} Post-Game Detection} (\cref{s:detection}).
\toolname performs an effective and efficient post-game detection on standard replay logs.
Note that \toolname requires no instrumentation beyond standard replay logging, incurring no additional recording or storage cost.
The detection pipeline has three stages:
(i) extracting honeypot-interaction aiming trajectories from tick-level logs via geometric gating and temporal clustering (\cref{s:interaction_extraction});
(ii) classifying each interaction based on its aiming pattern to identify abnormal aimbot-like trajectory (\cref{s:trajectory_identification});
and  (iii) aggregating interaction-level inferences into a player-level decision for the match (\cref{s:player_detection}).

\begin{figure*}[t]
    \centering
    \includegraphics[width=0.8\linewidth]{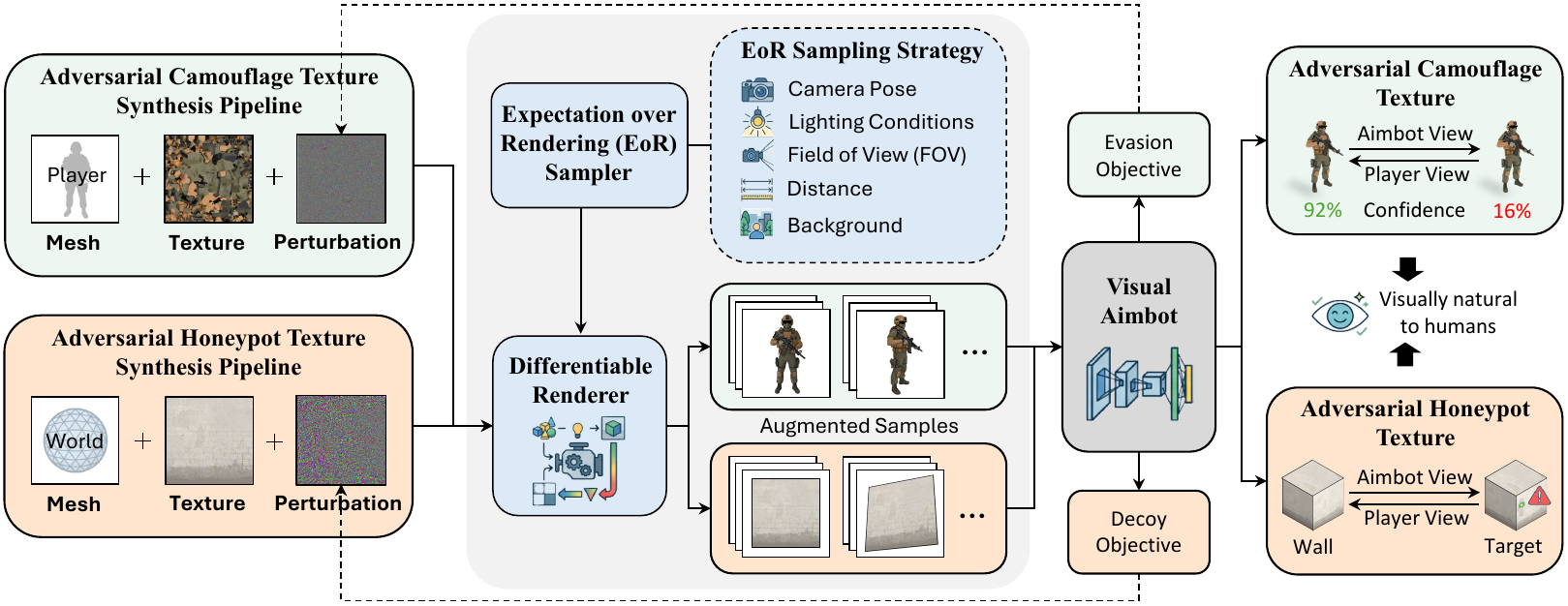}
    \caption{Adversarial texture synthesis.}\label{f:texture_synthesis}
\end{figure*}

\section{Adversarial Texture Synthesis}
\label{s:synthesis}

This section presents our adversarial texture synthesis pipeline (\cref{f:texture_synthesis}), covering preliminaries (\cref{s:preliminaries}), synthesis process (\cref{s:act}, \cref{s:aht}), and their secure and efficient management strategies (\cref{s:texture_management}, \cref{s:texture_deployment}).

\subsection{Preliminaries}
\label{s:preliminaries}

\noindent
\textbf{Expectation over Transformations (EoT).}
EoT~\cite{ICML18-eot,brown2017adversarial,xu2020adversarial,ruan2023towards} is a training-time strategy for crafting robust adversarial examples by optimizing not for a single input, but for the expected model loss under a distribution of transformations (e.g., scaling, rotation, viewpoint, lighting, noise, compression).

\noindent
\textbf{Expectation over Renderings (EoR).}
When adversarial examples must remain effective after a 3D rendering pipeline, nuisance variables arise from the rendering process that maps 3D assets (e.g., geometry, materials, and textures) and scene conditions (e.g., camera pose, distance, FOV, lighting, background, and motion blur) to the final image.
We therefore adapt EoT to \emph{Expectation over Renderings} (EoR), which optimizes asset-space parameters---here, the texture---so that the rendered outputs remain adversarial over a distribution of rendering states.
Let $M$ denote the mesh, $\tau$ the texture parameters to optimize, $\theta \sim D$ the rendering parameters, $R$ a differentiable renderer, $f$ the aimbot vision model, and $y$ the attack target (evasion or decoy). The EoR objective is
{\formulafont
\begin{equation}
\label{eq:eor}
\min_{\tau \in \mathcal{T}} \ \mathbb{E}_{\theta \sim D}\!\left[\mathcal{L}\big(f\big(R(M,\tau;\theta)\big), \, y\big)\right].
\end{equation}
}

\noindent
\textbf{Fully Differentiable Rendering Pipeline.}
To optimize under the EoR objective above, we integrate an end-to-end differentiable rendering pipeline~\cite{hull2024adversarial,pestana2022transferable}.
The renderer $R(M,\tau;\theta)$ maps the textured asset to an image $I$ given a render condition $\theta$.
Because rasterization and shading are differentiable in our setup, gradients of the detection loss with respect to the texture follow the chain rule:
{\formulafont
\begin{equation}
\label{eq:gradient_flow}
\tau \xrightarrow{\text{UV Mapping}} M \xrightarrow{R(\cdot;\,\theta)} I
\xrightarrow{\,f(\cdot)\,} \text{detections} \xrightarrow{\ \mathcal{L}\ } \nabla_{\tau}\mathcal{L},
\end{equation}
}i.e., $\nabla_{\tau}\mathcal{L} = \frac{\partial \mathcal{L}}{\partial f}\frac{\partial f}{\partial I}\frac{\partial R}{\partial \tau}$.

\subsection{Adversarial Camouflage Textures}
\label{s:act}

Adversarial Camouflage Textures (ACT) degrade aimbots' performance by reducing their confidence of game player models, thereby hindering localization and recognition.
To ensure robustness across rendering conditions, we optimize ACT under EoR framework.

\textit{Rendering Condition Distribution.}
Let the game-engine rendering condition be $\theta = (d,\phi,\psi,\lambda,b)$, where $d$ is camera distance ($d \in [d_{\min}, d_{\max}]$), $(\phi,\psi)$ are yaw/pitch angles ($\phi \in [\phi_{\min},\phi_{\max}], \ \psi \in [\psi_{\min},\psi_{\max}]$), $\lambda$ parameterizes lighting (e.g., ambient/point), and $b$ indexes background assets ($b \in \mathcal{B}$).
We define a distribution $\mathcal{D}$ over $\theta$ and, at each iteration $t$, draw a \emph{minibatch} of $m$ rendering states
{\formulafont\begin{equation}
\label{eq:rendering_distribution}
\Theta_t = \{\theta_1,\ldots,\theta_m\}, \qquad \theta_j \overset{\text{i.i.d.}}{\sim} \mathcal{D}.
\end{equation}
}

\textit{ACT's EoR Objective.}
We optimize ACT under \cref{eq:eor} using a stochastic estimator over the current minibatch $\Theta_t$:
{\formulafont
\begin{equation}
\label{eq:act_loss_stoch}
\widehat{\mathcal{L}}_{\mathrm{ACT}}(\tau;\Theta_t)
= \frac{1}{m} \sum_{j=1}^{m}
\ell_{\mathrm{det}}\!\big(f(R(M,\tau;\theta_j))\big),
\quad \theta_j \overset{\text{i.i.d.}}{\sim} \mathcal{D}.
\end{equation}
}This unbiased estimator is differentiated end-to-end through the rendering pipeline (\cref{eq:gradient_flow}) to obtain $\nabla_{\tau}\widehat{\mathcal{L}}_{\mathrm{ACT}}$.

\textit{Constrained Update.}
We update $\tau$ with projected gradient steps~\cite{pgd} in an $\ell_\infty$ ball of radius $\epsilon$ around the initial texture $\tau_0$, with per-channel clamping to $[0,1]$:
{\formulafont
\begin{align}
\label{eq:pgd_update}
\tilde{\tau}^{(t+1)} &= \tau^{(t)} - \alpha \,\mathrm{sign}\!\left(
\nabla_{\tau}\widehat{\mathcal{L}}_{\mathrm{ACT}}\big(\tau^{(t)};\Theta_t\big)\right),\\
\tau^{(t+1)} &= \mathrm{clip}_{[0,1]}\!\left(
\Pi_{\mathcal{B}_\infty(\tau_0,\epsilon)}\big(\tilde{\tau}^{(t+1)}\big)\right),
\end{align}
}where $\Pi_{\mathcal{B}_\infty(\tau_0,\epsilon)}(z) = \tau_0 + \mathrm{clip}_{[-\epsilon,\epsilon]}(z-\tau_0)$ is the projection onto the $\ell_\infty$ ball about $\tau_0$.
At each iteration, we resample $\Theta_t$ (viewpoints, distances, lighting, backgrounds), accumulate minibatch gradients via \cref{eq:act_loss_stoch}, and apply \cref{eq:pgd_update}.

\subsection{Adversarial Honeypot Textures}
\label{s:aht}

Adversarial Honeypot Textures (AHT) are embedded in 3D world to \emph{expose} aimbot behavior: they are visually natural to human players, yet appear to vision-based aimbots as highly salient, target-like objects.
We again use the EoR strategy to ensure robustness under rendering variability.

\textit{Rendering Condition Distribution.}
We follow nearly the same rendering distribution $\mathcal{D}$ as in \cref{eq:rendering_distribution}, excluding the background $b$ term since the honeypot is part of the scene.

\textit{Targeted Detection with Uniqueness.}
We want a \emph{single} high-confidence decoy in aimbot's detection area, suppressing competitors; otherwise multiple decoys can affect aimbot's decision-making process, leading to unstable aiming behavior.
Let $p_1$ be the top detection score and $p_2$ the highest-confidence \emph{non-overlapping} competitor.
We enforce this with a margin and confidence calibration term:
{\formulafont
\begin{equation}
\label{eq:aht_rank}
\mathcal{L}_{\mathrm{rank}}
= a \,\mathrm{ReLU}\,\big(m - (p_1 - p_2)\big) + b\, (1 - p_1)^2,
\end{equation}
}where $m$ is the desired margin and $a,b$ weight the terms.

\textit{Geometric Consistency.}
To reduce unstable locks under viewpoint changes, we regularize the top box toward desired normalized dimensions.
Let $(w,h)$ be the width/height of the top box, and $(W_T,H_T)$ be the reference extents for normalization under $\theta_j$.
Define $w_{\mathrm{frac}} = w/W_T$, $h_{\mathrm{frac}} = h/H_T$ and targets $w_{\mathrm{frac}}^{*}, h_{\mathrm{frac}}^{*}$.
The geometric loss is
{\formulafont
\begin{equation}
\label{eq:aht_geo}
\mathcal{L}_{\mathrm{geo}} = \mu \,\big|w_{\mathrm{frac}} - w_{\mathrm{frac}}^{*}\big|
+ \nu \,\big|h_{\mathrm{frac}} - h_{\mathrm{frac}}^{*}\big|,
\end{equation}
}with $\mu,\nu>0$ weighting width/height deviations.

\textit{AHT's EoR Objective.}
We optimize AHT using a stochastic estimator over $\Theta_t$:
{\formulafont
\begin{equation}
\label{eq:aht_loss_stoch}
\widehat{\mathcal{L}}_{\mathrm{AHT}}(\tau;\Theta_t)
= \frac{1}{m} \sum_{j=1}^{m}
\Big[\lambda_{r}\,\mathcal{L}_{\mathrm{rank}}^{(j)} + \lambda_{g}\,\mathcal{L}_{\mathrm{geo}}^{(j)}\Big],
\quad \theta_j \overset{\text{i.i.d.}}{\sim} \mathcal{D},
\end{equation}
}which is differentiated end-to-end through the pipeline (\cref{eq:gradient_flow}) to obtain $\nabla_{\tau}\widehat{\mathcal{L}}_{\mathrm{AHT}}$.
Finally, we update the texture $\tau$ following the same method described in \cref{eq:pgd_update}.

\textit{Cross-Model Transferability Optimization.}
We enhance the cross-model transferability of AHTs by combining proxy-model ensembling (ENS), momentum iterative optimization (MI), and translation-invariant optimization (TI).
At each EoR-PGD step, ENS averages the attack loss across an ensemble of proxy models.
TI then smooths the resulting texture gradient using a $7\times7$ Gaussian kernel, reducing sensitivity to spatial shifts.
Finally, MI accumulates the $\ell_1$-normalized smoothed gradient with momentum factor $\mu=1.0$.
The sign of the accumulated gradient is used in the projected update (\cref{eq:pgd_update}), encouraging the optimization to identify perturbation directions shared across model decision boundaries and thereby improving cross-model transferability (\cref{e:transferability}).

\subsection{Secure Adversarial Texture Management}
\label{s:texture_management}

This section details the security guarantees of adversarial textures, enforced by the following three mechanisms.

\noindent\textbf{Randomized Synthesis and Per-Session Selection.}
Following well-established practices to counter predictability, such as address space layout randomization (ASLR)~\cite{shacham2004aslr}, we randomize both texture synthesis and per-game-session selection.
In detail, we synthesize a \emph{diverse} pool of adversarial textures offline by varying the synthesis initialization (e.g., random seeds).
Then we \emph{randomly} sample a subset from the pool for each game session (\circled{1}, \cref{f:overview}).
Such randomization obscures both which textures are adversarial and the specific perturbation patterns in each texture, limiting an aimbot developer's ability to anticipate or adapt across sessions.

\noindent\textbf{In-Memory-Only Deployment.}
By default, CS2 stores normal textures on disk, and our design preserves this behavior for normal assets.
For \emph{adversarial} textures, however, we depart from disk storage: they are transmitted over the network from the server, kept solely in the client's memory, and then cleared immediately in memory when the match ends (\circled{2}, \circled{4}, \cref{f:overview}).
This in-memory-only deployment keeps the adversarial textures within the protection scope of existing memory-based anti-cheat mechanisms (see Assumptions in \cref{s:problem_statement}).
By avoiding any persistent disk storage for these textures, it prevents an aimbot developer's ability from extracting, reverse-engineering, or adapting to our textures.

\noindent\textbf{Asset Integrity Verification.}
Because our textures are applied to existing client-side 3D models, an adversary might attempt to degrade effectiveness by altering models or UV mappings~\cite{uv_mapping}.
To prevent such tampering, we verify the integrity of all game assets loaded from the client's disk (\circled{2}, \cref{f:overview}).
This integrity verification follows standard practice already enforced by vanilla clients (e.g., Valve's VPK and Unreal Engine's PAK file integrity check).
Thus, our approach introduces negligible integration overhead.

\subsection{Efficient Adversarial Texture Deployment}
\label{s:texture_deployment}

This section explains how we achieve efficient texture deployment through the following three design choices.

\noindent 
\textbf{Offline Synthesis}.
We pre-synthesize a diverse pool of adversarial textures on the server, entirely \emph{offline}, so no on-the-fly synthesis is incurred during live gameplay.
The average synthesis time for AHT is 12\,s and ACT 75\,s (\cref{e:overhead}).

\noindent
\textbf{Cross-Session Reuse}.
The server reuses each texture across multiple sessions to amortize the synthesis cost.
As reported by \cite{cs2-online-player}, CS2 has averaged \mbox{902k} online players over the past two years, peaking at \mbox{1.8M}.
Assuming 10 players per match, this corresponds to roughly \mbox{90k} concurrent matches on average.
At this scale, reusing each texture for hundreds to thousands of sessions still yields ample variety while keeping costs manageable.
Suppose each texture is reused for 1{,}000 sessions, the amortized per-session cost is \mbox{$\approx$0.012\,s} (AHT) and \mbox{$\approx$0.075\,s} (ACT), both $<0.1$\,s.
Such overheads are readily affordable for production servers.

\noindent
\textbf{Lightweight Network Transmission}.
Adversarial textures are lightweight image files, far smaller than heavyweight assets, such as 3D models.
In our implementation, AHTs average \mbox{0.83\,MB} and ACTs \mbox{7.48\,MB} (\cref{e:overhead}).
Thus, per-session download introduces negligible bandwidth and latency overhead to the whole system.

\section{Aimbot Cheating Detection}
\label{s:detection}

\begin{figure}[t!]
    \centering
    \includegraphics[width=1\linewidth]{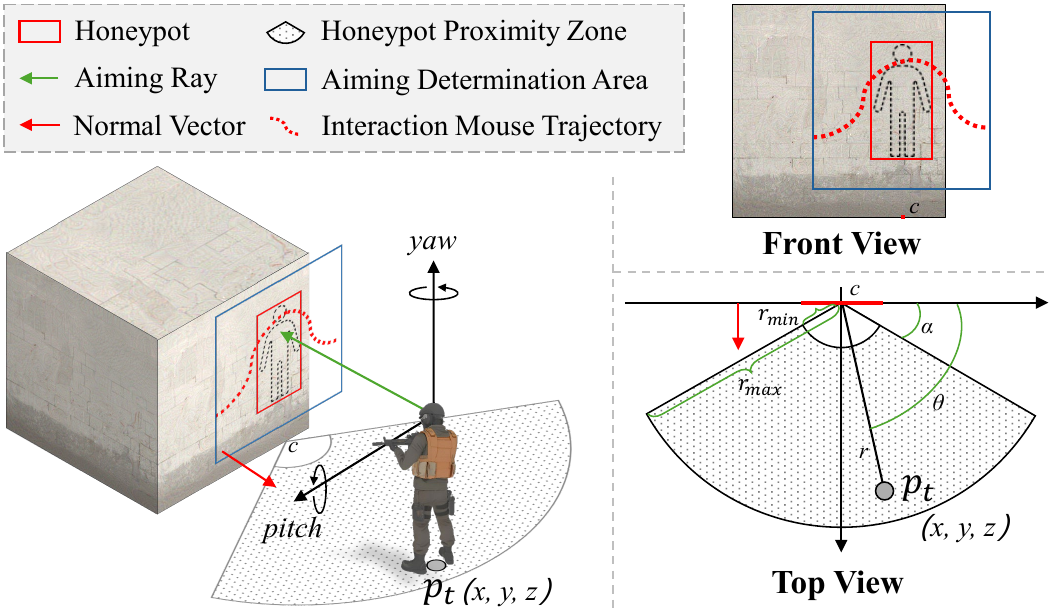}
    \caption{Adversarial honeypot for aimbot detection}\label{f:detection}
\end{figure}

This section presents our post-game aimbot detection, comprising three phases: (i) honeypot interaction trajectory extraction (\cref{s:interaction_extraction}), (ii) abnormal trajectory identification (\cref{s:trajectory_identification}), and (iii) player-level cheating detection (\cref{s:player_detection}).

\subsection{Honeypot-Interaction Mouse-Aiming Trajectory Extraction}
\label{s:interaction_extraction}

\noindent
\textbf{Extraction Overview.}
We convert tick-level replay logs (detailed in \cref{a:replay}) into precise honeypot-interaction mouse aiming trajectories in two stages.
First, \emph{geometric gating} retains only spatially-valid ticks: the player must be within the \emph{Honeypot Proximity Zone} and their view (aiming) ray must intersect the \emph{Aiming Determination Area} on the honeypot plane, as illustrated in \cref{f:detection}.
Second, \emph{temporal clustering} groups valid ticks to segment contiguous interactions.
For each cluster, the intersections between the aiming ray and the honeypot plane define a 2D honeypot-interaction mouse trajectory.

\noindent\textbf{Honeypot Proximity Zone (HPZ).}
We define HPZ as the neighborhood around each honeypot in which a player is considered spatially eligible for interaction.
Let $c$ be the honeypot center projected to the floor plane, $p_t$ the player's floor-projected position at tick $t$.
Define $r=\lVert p_t - c\rVert$ and $\theta$ the incidence angle relative to the honeypot surface.
We instantiate the HPZ as the circular sector, as illustrated in the top view of \cref{f:detection}:
{\formulafont
\[
\mathrm{HPZ}\;=\;\bigl\{(r,\theta)\ \big|\ r_{\min}\le r\le r_{\max},\ \alpha\le \theta \le \pi-\alpha \bigr\},
\]
}excluding viewpoints too close, far, or excessively grazing.
Ticks with $(r,\theta)\notin\mathrm{HPZ}$ are discarded to improve accuracy and efficiency.

\noindent\textbf{Aiming Determination Area (ADA).}
Conditioned on being in HPZ, we further require that the player's view ray intersects a target region on the honeypot plane (see the front view in \cref{f:detection}).
The ADA is an axis-aligned 2D box on the honeypot plane, centered at the honeypot.
A tick is \emph{aim-valid} if the ray cast from the player's position along their (\textit{Yaw}, \textit{Pitch}) direction intersects this box.
The resulting aiming trajectory is the ordered sequence of 2D intersection points within ADA.
We size ADA slightly larger than the physical honeypot to capture enough aiming patterns.

\noindent\textbf{Temporal Clustering.}
After HPZ/ADA gating, we cluster the remaining tick indices using 1-D DBSCAN to form contiguous interaction segments.
We set the neighborhood radius to $\varepsilon=64$ ticks ($\approx 1.0$\,s) and the minimum samples to $m=32$ ticks ($\approx 0.5$\,s), so that clusters correspond to sustained interactions rather than spurious single-frame hits.
Each cluster yields a 2D mouse trajectory in ADA plane, representing a complete honeypot interaction.

\subsection{Abnormal Interaction Identification}
\label{s:trajectory_identification}

Given the extracted honeypot-interaction aiming trajectories, we formulate abnormal trajectory detection as supervised sequence classification.
Each trajectory is labeled and encoded with temporal deltas, 2D intersection points, and honeypot context, and a bidirectional LSTM is trained to distinguish abnormal from normal trajectories.
Below, we describe our approach in detail.

\textit{Dataset.}
We build a dataset of 3{,}074 labeled trajectories for training and evaluating our classification model (\cref{e:trajectory_identification}).

\textit{Sequence representation and normalization.}
Given a clustered mouse aiming trajectory of length $T$, we encode each tick into a four-dimensional token
{\formulafont
\[
s_{t} = [\Delta t, x_{t}, y_{t}, 1\{(x_{t}, y_{t}) \in \mathrm{HP}\}] \in \mathbb{R}^{4}\quad (t=1\ldots T),
\]
}where $\Delta t = (tick_{t} - tick_{t-1}) / 64$ normalized tick rate, $(x_{t}, y_{t})$ are 2D coordinates in ADA, $[0, 1]$ indicates whether the hit is inside the honeypot $\mathrm{HP}$.
In addition, we append the honeypot's normalized box coordinates $[x_{\mathrm{HP}}, y_{\mathrm{HP}}, w_{\mathrm{HP}}, h_{\mathrm{HP}}]$ as static features.
Sequences are truncated or padded to $T_{\max} = 320$ with masks.

\textit{Model architecture.}
We encode each sequence with a bidirectional LSTM.
Let the final forward and backward states be $h^{\rightarrow}, h^{\leftarrow} \in \mathbb{R}^{d}$ and define the sequence embedding $h = [h^{\rightarrow} || h^{\leftarrow}] \in \mathbb{R}^{2d}$. We concatenate $h$ with $b_{\mathrm{HP}}$ and feed it to a two-layer MLP $g(\cdot)$ with BatchNorm and Dropout: $z = g([h || b_{\mathrm{HP}}]) \in \mathbb{R}$, where $z$ is the logit for the positive class.
We train at the sequence level using class-weighted binary cross-entropy with logits and $\ell_2$ regularization.
For a batch $\{(S_i, b_{\mathrm{HP},i}, y_i)\}_{i=1}^{N}$ with $y_i \in \{0,1\}$ and logits $z_i = g([h_i || b_{\mathrm{HP},i}])$,
{\formulafont
\[
BCELogit(z_i, y_i) = \log(1 + e^{z_i}) - y_i z_i,
\]
\[
\mathcal{L}_{\text{data}} = \frac{1}{N} \sum_{i=1}^{N} w_{y_i} \, BCELogit(z_i, y_i),
\]
}where $w_1 = \rho$, $w_0 = 1$, and $\rho = N_{\text{neg}} / N_{\text{pos}}$ when positives are rarer.
The final objective is
{\formulafont
\[
\mathcal{L} = \mathcal{L}_{\text{data}} + \lambda \|\theta\|_2^2,
\]
}with $\theta$ denoting all trainable parameters and $\lambda$ set via weight decay in the optimizer.

\subsection{Player-Level Aimbot Cheating Detection}
\label{s:player_detection}

This section formalizes our player-level decision rules based on trajectory-level identification.

\noindent\textbf{Cheating Decision Making.}
Trajectory-level identification (\cref{s:trajectory_identification}) does not directly translate into player-level cheating detection.
Instead, we aggregate predictions across multiple honeypot-interaction trajectories into a \emph{player-level decision}: a player is flagged if they exhibit at least $T$ abnormal honeypot interactions in a match.

\noindent\textbf{Player-Level Cheating Detection.}
We model each honeypot interaction within a match as independent, as honeypots are placed at distinct locations and interactions occur at different times.
Let $\mathrm{TPR}_T$ and $\mathrm{FPR}_T$ denote per-trajectory true-positive and false-positive rate respectively.
For a match with $n$ interaction opportunities, the number of abnormal interactions follows a Binomial distribution, and a player is flagged when at least $T$ of the $n$ interactions are classified abnormal.
The resulting player-level true-positive rate $\mathrm{TPR}_P$ and false-positive rate $\mathrm{FPR}_P$ are the corresponding binomial tails:

{\formulafont
\[
\mathrm{TPR}_P(n,T)=\sum_{k=T}^{n}\binom{n}{k}\,\mathrm{TPR}_T^{\,k}\,(1-\mathrm{TPR}_T)^{n-k},
\]
\[
\mathrm{FPR}_P(n,T)=\sum_{k=T}^{n}\binom{n}{k}\,\mathrm{FPR}_T^{\,k}\,(1-\mathrm{FPR}_T)^{n-k}.
\]
}

\noindent\textbf{Tunable Thresholds.}
Developers can tune the aggregation threshold $T$ to trade recall for precision: increasing $T$ lowers $\mathrm{FPR}_P$ at the cost of $\mathrm{TPR}_P$, and decreasing $T$ does the opposite (\cref{t:player-level_detection} in \cref{e:trajectory_identification}; \cref{a:nablation} details the dependence on the interaction count $n$).
Because mis-flagging legitimate players undermines the anti-cheat system's \emph{trustworthiness}, real-world deployments typically prioritize a low false-positive rate over occasional misses.
In our real-game evaluation (\cref{e:real-world_detection}), we empirically find thresholds $T \in [3,5]$ provide a practical operating range, accurately identifying player-level cheating while keeping no false alarm.

\section{Evaluation}
\label{s:evaluation}
We answer the following research questions (RQs):

\begin{itemize}
    \item \textbf{RQ1.} How effective is \toolname in proactively defending against vision cheating during gameplay (\cref{e:defense_effectiveness})?
    \item \textbf{RQ2.} How does our detection model perform in distinguishing cheating honeypot-interaction trajectories from normal trajectories?
    How does it perform at player-level?
    Does the visualized mouse trajectory reveal a strong signal?
    (\cref{e:trajectory_identification})
    \item \textbf{RQ3.} How effective is our detection model in real-game scenarios, and how does its performance compare against state-of-the-art baselines? (\cref{e:real-world_detection})
    \item \textbf{RQ4.} How robust is our approach under diverse game rendering conditions (\cref{e:robustness})?
    \item \textbf{RQ5.} What is \toolname's performance overhead (\cref{e:overhead})?
    \item \textbf{RQ6.} Do adversarial textures degrade real-world player experience (e.g., distraction or visual strain) (\cref{e:gaming_impact})?
    \item \textbf{RQ7.} How well does our approach transfer across different aimbot models (\cref{e:transferability})?
\end{itemize}

\subsection{Experimental Setup}
\label{s:experimental_setup}

\noindent\textbf{Implementation.}
We provide implementation details in \cref{a:implementation}.

\noindent \textbf{Texture Assets.}
We evaluate scalability by synthesizing adversarial textures from real-game assets and 3D generation pipelines~\cite{tripo,meshy,hyer3d}.
\emph{AHT} uses 24 official CS2 textures across 7 material types.
\emph{ACT} uses Tripo~\cite{tripo} to generate 4 male and female player models (\cref{a:tripo}).

\noindent \textbf{Map Construction and Honeypot Deployment.}
We use Valve's Hammer editor~\cite{hammer} to build a custom CS2 map, whose structure is inspired by the classic CS1.6 map, \texttt{iceworld} (detailed in \cref{a:implementation}).
This map deploys 16 AHTs at diverse locations.

\noindent \textbf{Aimbot and Proxy Models.}
We evaluate a widely used open-source YOLOv5-based aimbot~\cite{rootkit_aimbot} (2.1k GitHub stars) at its default confidence threshold of $0.4$.
Its default mouse smoothing applies $40\%$ of the target offset per step, producing human-like trajectories instead of instant snaps (see \cref{a:aimbot}).
Following prior work~\cite{cloak}, we optimize against \emph{YOLOv5n} as the proxy and primarily evaluate on \emph{YOLOv5s} as the attacker.
We evaluate AHT transfer boosters across YOLOv5/8, Faster R-CNN, and NanoDet (\cref{e:transferability}).

\subsubsection{Evaluation Metrics}
\label{s:evaluation_metrics}

We assess \toolname using the following four metrics.

\textbf{Decoy Success Rate (DSR).}
Measures how effectively AHTs \emph{attract} the aimbot.
We consider a successful deception when aimbot's top-1 detection is a honeypot with confidence $\ge0.4$ (default threshold~\cref{s:experimental_setup}).
We evaluate DSR across various conditions, including different camera distances and angles (\cref{e:robustness}).

\textbf{Evasion Success Rate (ESR).}
Measures how effectively ACTs \emph{hide} players from the aimbot.
We consider a successful evasion if no player-aligned detection exceeds $0.4$.
ESR is reported under the same rendering variations.

\textbf{Uniqueness Rate (UR).}
Measures whether AHT yields a \emph{single}, unambiguous attractor (\cref{s:aht}).
We consider it unique if exactly \emph{one} honeypot exceeds detection threshold.
High uniqueness reduces multi-target ambiguity for both attraction and detection.

\textbf{Structural Similarity Index (SSIM)}~\cite{wang2004image}.
Quantifies perceptual similarity between original textures and adversarial textures (AHT/ACT) on RGB maps.
We report mean SSIM in $[0,1]$, where higher values indicate closer resemblance to the original asset.

\subsection{Proactive Defense Effectiveness}
\label{e:defense_effectiveness}

\begin{table}[t]
\centering
\begin{talltblr}[
    caption = {Effectiveness of Adversarial Camouflage Textures (ACT) for evading aimbots},
    label = {t:effectiveness_act}
] {
    stretch = 0.5,
    colspec = {c|cc|cc},
    rows = {font=\footnotesize},
}
    \hline[1pt]
    \textbf{} &
    \SetCell[c=2]{c} \textbi{Male Character} & & 
    \SetCell[c=2]{c} \textbi{Female Character} & \\ 
    \hline
    \SetCell[r=2]{m} {\textbf{Original}\\\textbf{Texture}\\\textbf{Example}} &
    Texture & Aimbot & 
    Texture & Aimbot \\
    &
    \includegraphics[width=1.2cm]{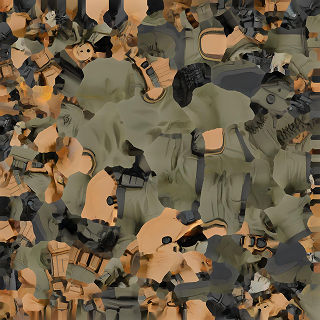} &  
    \includegraphics[width=1.2cm]{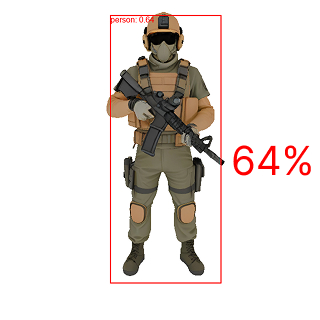} &  
    \includegraphics[width=1.2cm]{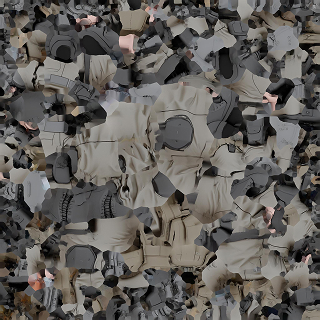} &  
    \includegraphics[width=1.2cm]{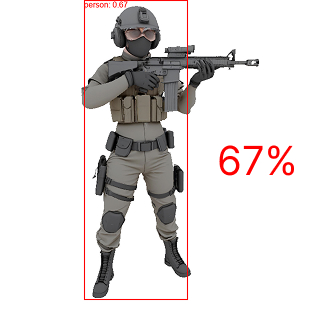} \\
    \hline
    \SetCell[r=2]{m} {\textbf{Adversarial}\\\textbf{Camouflage}\\\textbf{Texture}\\\textbf{Example}} &
    Texture & Aimbot &
    Texture & Aimbot \\
    &
    \includegraphics[width=1.2cm]{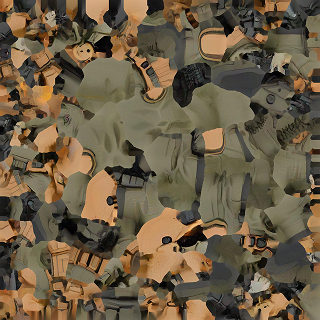} &  
    \includegraphics[width=1.2cm]{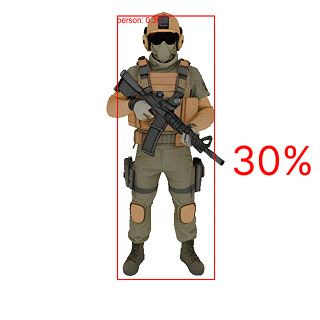} &  
    \includegraphics[width=1.2cm]{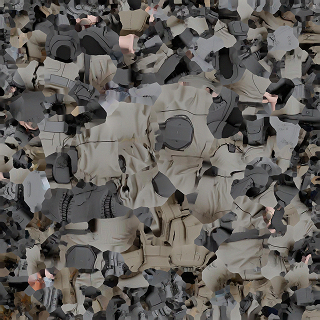} &  
    \includegraphics[width=1.2cm]{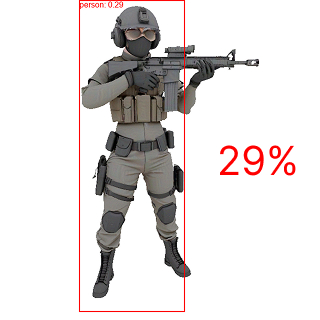} \\
    \hline[1pt]
\end{talltblr}

\vspace{1pt}

\begin{tblr} {
    stretch = 0.5,
    colspec = {c|X[1.16cm,c]|X[1.16cm,c]|X[1.16cm,c]},
    rows = {font=\footnotesize},
    row{1} = {font=\small}
}
    \hline[1pt]
    \textbf{} & \textbi{Male} & \textbi{Female} & \textbi{Total} \\ 
    \hline
    \textbf{\# of Textures} & 2 & 2 & 4 \\
    \hline
    \textbf{\# of ACT} & 20 & 20 & 40 \\
    \hline
    \textbf{\# of EoR Sampling} & 25,600 & 25,600 & 51,200 \\
    \hline
    \textbf{Original Confidence Score} & 67.7\% & 56.8\% & 62.3\% \\
    \hline
    \textbf{ACT Confidence Drop} & $\downarrow 49.9\%$ & $\downarrow 42.5\%$ & $\downarrow 46.3\%$ \\
    \hline
    \textbf{ACT Confidence Score} & \textbf{17.8\%} & \textbf{14.3\%} & \textbf{16.0\%} \\
    \hline
    \textbf{SSIM} & 81.5\% & 80.5\% & 81.0\% \\
    \hline
    \textbf{Evasion Success Rate} & \textbf{84.1\%} & \textbf{86.0\%} & \textbf{85.1\%} \\
    \hline[1pt]
\end{tblr}
\end{table}

\begin{table*}[t]
\centering
\begin{talltblr}[
    caption = {Effectiveness of Adversarial Honeypot Textures (AHT) for misleading aimbots},
    label = {t:effectiveness_aht},
] {
    stretch = 0.5,
    colspec = {X[2.6cm,c]|c|c|c|c|c|c|c|X[1.3cm,c]},
    rows = {font=\footnotesize},
    rows = {valign=b}
}
    \hline[1pt]
    \textbf{Material} & \textbi{Brick} & \textbi{Cinderblock} & \textbi{Concrete} & \textbi{Mudbrick} & \textbi{Plaster} & \textbi{Sand} & \textbi{Stone} & \textbi{Total} \\
    \hline
    \textbf{Original \\ Texture \\ Example \\ } & 
    \includegraphics[width=1.3cm]{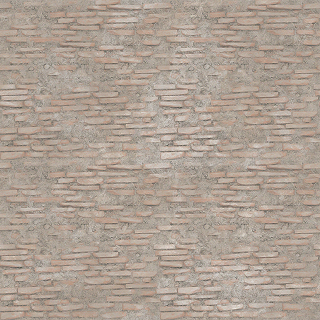} &  
    \includegraphics[width=1.3cm]{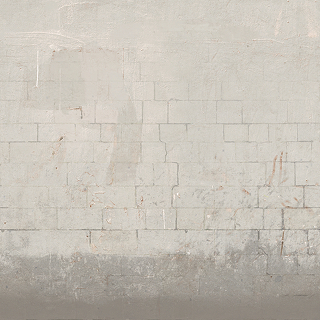} &  
    \includegraphics[width=1.3cm]{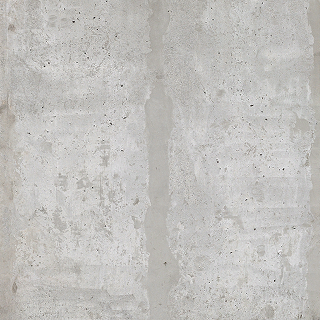} &  
    \includegraphics[width=1.3cm]{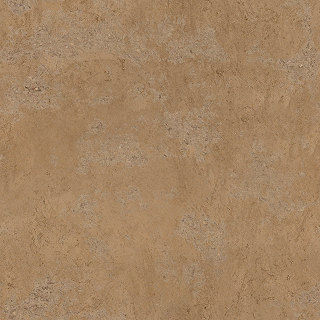} &  
    \includegraphics[width=1.3cm]{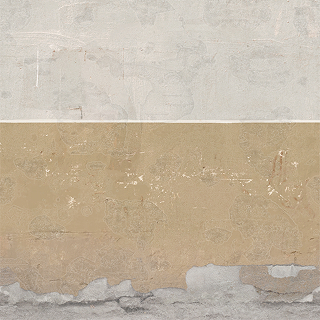} &  
    \includegraphics[width=1.3cm]{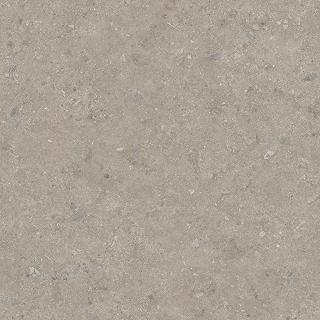} &  
    \includegraphics[width=1.3cm]{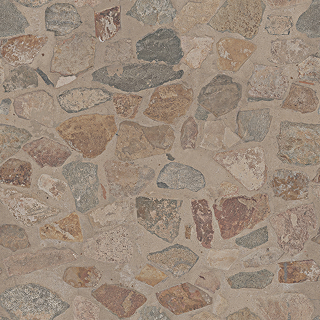} &  
    \SetCell[r=2]{c} \includegraphics[width=1.0cm]{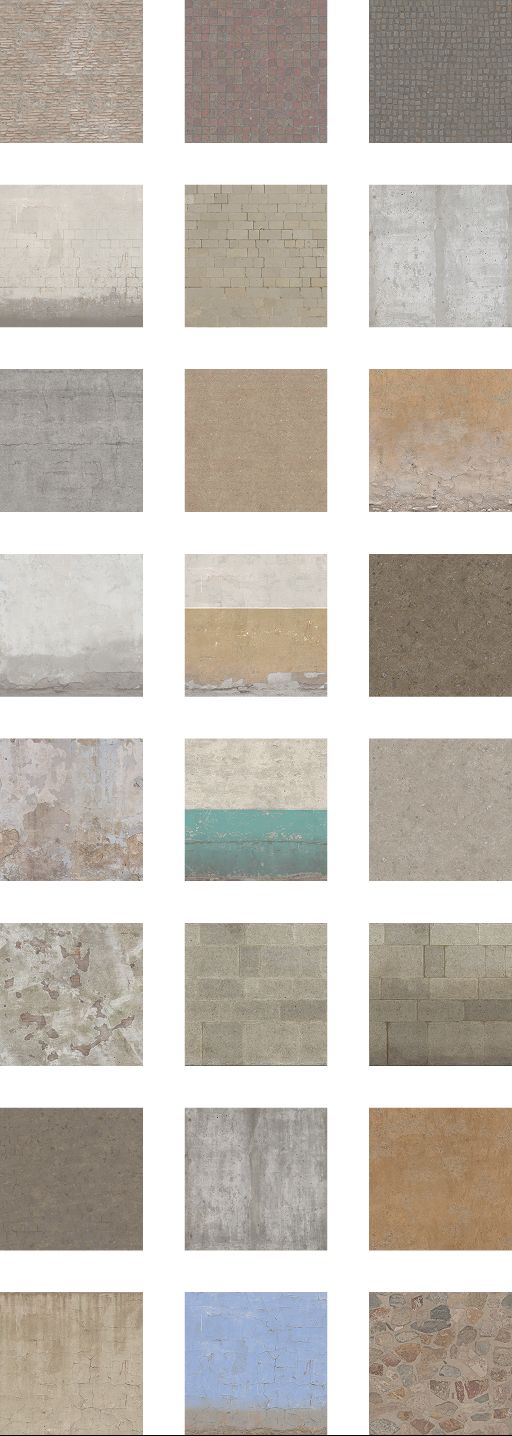}   \\
    \hline
    \textbf{Adversarial \\ Honeypot \\ Texture \\ Example \\} &
    \includegraphics[width=1.3cm]{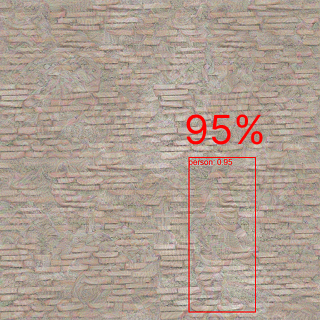} &  
    \includegraphics[width=1.3cm]{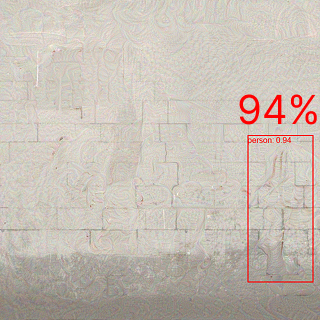} &  
    \includegraphics[width=1.3cm]{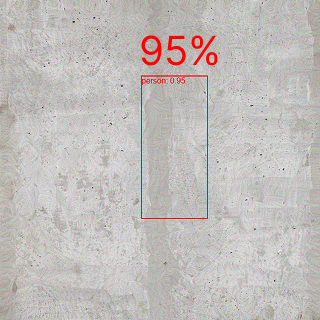} &  
    \includegraphics[width=1.3cm]{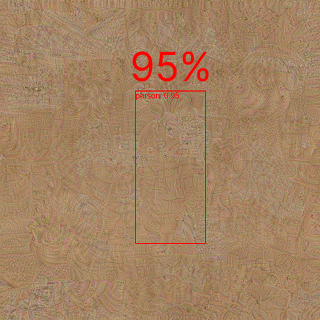} &  
    \includegraphics[width=1.3cm]{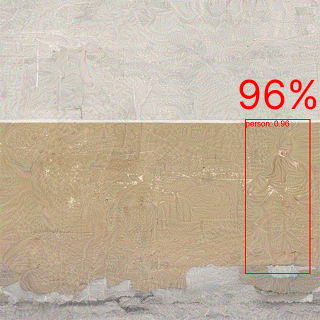} &  
    \includegraphics[width=1.3cm]{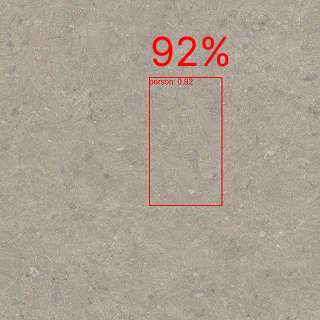} &  
    \includegraphics[width=1.3cm]{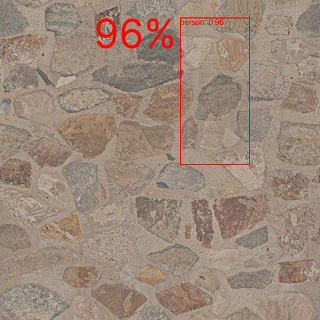} &
    \\
    \hline
    \textbf{\# of Textures} & 4 & 2 & 3 & 3 & 6 & 2 & 4 & 24 \\
    \hline
    \textbf{\# of AHT} & 40 & 20 & 30 & 30 & 60 & 20 & 40 & 240 \\
    \hline
    \textbf{\# of EoR Sampling} & 20,480 & 10,240 & 15,360 & 15,360 & 30,720 & 10,240 & 20,480 & 122,880 \\
    \hline
    \textbf{Confidence Score} &
    82.5\% & 75.4\% & 72.7\% & 72.1\% & 79.3\% & 87.8\% & 84.2\% & 79.3\% \\
    \hline
    \textbf{Uniqueness Rate} &
    98.3\% & 99.3\% & 93.3\% & 95.3\% & 94.7\% & 100.0\% & 99.8\% & 96.9\% \\
    \hline
    \textbf{SSIM} &
    70.9\% & 63.7\% & 72.5\% & 70.4\% & 67.3\% & 75.3\% & 77.3\% & 71.0\% \\
    \hline
    \textbf{Decoy Success Rate} & 
    \textbf{98.3\%} & \textbf{99.3\%} & \textbf{93.3\%} & \textbf{95.6\%} & \textbf{94.7\%} & \textbf{100.0\%} & \textbf{99.9\%} & \textbf{96.9\%} \\
    \hline[1pt]
\end{talltblr}
\end{table*}

To answer \textbf{RQ1}, we evaluate the two defenses provided by \toolname:
(i) Adversarial Camouflage Textures (ACT) for \emph{evasion} (\cref{t:effectiveness_act}), and
(ii) Adversarial Honeypot Textures (AHT) for \emph{deception} (\cref{t:effectiveness_aht}).

\noindent \textbf{ACT (Evasion).}
We assess ACTs using Evasion Success Rate (ESR), the aimbot's confidence score, and visual fidelity (SSIM).
We consider four distinct player models (covering both male and female characters), and for each model we synthesize ten ACT variants, yielding 40 ACTs in total.
To ensure robustness, we adopt Expectation-over-Rendering optimization, randomly sampling 1{,}280 combinations of various conditions (detailed in \cref{e:robustness}) per ACT (51{,}200 EoR samples overall).
We evaluate the corresponding \emph{original} textures under identical conditions as a baseline.

\cref{t:effectiveness_act} shows that ACTs substantially depress the aimbot's confidence in detecting the player, reducing mean confidence from 62.3\% to \textbf{16.0\%} (a \emph{46.3\%} drop), well below the default detection threshold $40\%$.
Concurrently, ACTs maintain high perceptual quality with a mean SSIM of 81.0\%, indicating strong visual fidelity.
The average ESR is \textbf{85.1\%} across rendering conditions, demonstrating effective proactive evasion against vision-based aimbots.

\noindent \textbf{AHT (Deception).}
We measure AHTs using Decoy Success Rate (DSR), Uniqueness Rate, aimbot confidence on honeypots, and SSIM.
Following \cref{s:experimental_setup}, we curate 24 real-game textures spanning seven material types (e.g., brick, cinderblock) and synthesize ten AHT variants per texture, producing 240 AHTs in total.
For each AHT, we sample 512 EoR condition combinations (122{,}880 samples overall); condition details are provided in \cref{e:robustness}.

AHTs reliably draw the aimbot's top-1 detection to the honeypot with high confidence (mean 79.3\%), yielding a \textbf{DSR of 96.9\%}.
Moreover, AHTs exhibit a \textbf{Uniqueness Rate of 96.9\%}, indicating that, in almost all views, exactly \emph{one} honeypot surpasses the threshold---providing a clean, unambiguous signal for downstream cheat detection.
Perceptual quality remains acceptable (mean SSIM 71.0\%), enabling AHTs to blend into the scene without distracting human players (real player impact study is in \cref{e:gaming_impact}).

\subsection{Cheating Honeypot Interaction Identification Performance}
\label{e:trajectory_identification}

\begin{table}[h]
\centering
\begin{talltblr} [
    caption = {\toolname's performance on honeypot-interaction cheating trajectory identification},
    label = {t:trajectory_identification}
]{
    stretch = 0.5,
    rows = {font=\small},
    colspec = {c|c|c|c|c},
}
\hline[1pt]
\textbf{Accuracy} & \textbf{Precision} & \textbf{Recall} & \textbf{F1} & \textbf{AUC-ROC} \\
\hline
98.05\%  & 96.53\%    & 98.98\% & 97.74\% & 99.29\% \\
\hline[1pt]
\end{talltblr}
\end{table}

\noindent \textbf{Setup.}
Following \cref{s:detection}, we set the honeypot proximity area parameters to $r_{min}=60.0$, $r_{max}=520.0$, and $\alpha=30.0$, and the aiming determination area to $256\times256$ pixels.

\noindent \textbf{Dataset.}
We record honeypot interactions extracted from replay logs on our custom map (\cref{s:experimental_setup}) from two controlled sources: human-honeypot interaction and aimbot-honeypot interaction (configured in \cref{a:aimbot}).
Labels are therefore determined by the recording condition rather than by manual annotation: trajectories from aimbot-assisted interactions are labeled \emph{aimbot} (positive), and those from human-only interactions are labeled \emph{human} (negative).
From these logs, we extract honeypot-interaction trajectories with the pipeline of \cref{s:interaction_extraction}, i.e., HPZ/ADA geometric gating followed by temporal clustering, yielding 3{,}074 aiming trajectories in total: 1{,}326 \emph{aimbot} and 1{,}748 \emph{human}.
We split the data into 2{,}151/461/462 sequences for training/validation/testing.

\noindent\textbf{Trajectory-Level Detection Results.}
\cref{t:trajectory_identification} summarizes \toolname's performance on trajectory-level cheating detection: accuracy \textbf{98.05\%}, precision \textbf{96.53\%}, recall \textbf{98.98\%}, and F1 \textbf{97.74\%}, with \textbf{99.29\%} AUC-ROC indicating excellent separability.
Errors are rare: a small number of false alarms (7/462), while missed cheating trajectories are extremely uncommon (2/462).
Consistently, the trajectory-level false positive rate is \textbf{2.64\%} and the false negative rate is \textbf{1.02\%},
supporting the model's suitability for high-recall screening of abnormal honeypot-interaction trajectories.

\begin{table}[h]
\centering
\begin{talltblr}[
    caption = {Player-level cheating detection performance.},
    remark{Note} = {$\mathrm{TPR}_P$ in \%. We report the compact case $n{=}T$; \cref{a:nablation} evaluates other $n$. A $k$-match decision flags a player only when flagged in all $k$ matches.},
    label = {t:player-level_detection}
] {
    stretch = 0.5,
    colspec = {c|cc|cc|cc},
    rows = {font=\scriptsize},
}
\hline[1pt]
\SetCell[r=2]{c} {\boldmath$T$}
& \SetCell[c=2]{c} {\bf One-Match} & & \SetCell[c=2]{c} {\bf Two-Match} & & \SetCell[c=2]{c} {\bf Three-Match} \\
\hline
& {\boldmath$\mathrm{TPR}_P$} & {\boldmath$\mathrm{FPR}_P$} & {\boldmath$\mathrm{TPR}_P$} & {\boldmath$\mathrm{FPR}_P$} & {\boldmath$\mathrm{TPR}_P$} & {\boldmath$\mathrm{FPR}_P$} \\
\hline
1 & $99.0$ & $2.6\!\times\!10^{-2}$ & $98.0$ & $7.0\!\times\!10^{-4}$ & $97.0$ & $1.8\!\times\!10^{-5}$ \\
2 & $98.0$ & $7.0\!\times\!10^{-4}$ & $96.0$ & $4.9\!\times\!10^{-7}$ & $94.0$ & $3.4\!\times\!10^{-10}$ \\
3 & $97.0$ & $1.8\!\times\!10^{-5}$ & $94.0$ & $3.4\!\times\!10^{-10}$ & $91.2$ & $6.2\!\times\!10^{-15}$ \\
4 & $96.0$ & $4.9\!\times\!10^{-7}$ & $92.1$ & $2.4\!\times\!10^{-13}$ & $88.4$ & $1.1\!\times\!10^{-19}$ \\
5 & $95.0$ & $1.3\!\times\!10^{-8}$ & $90.3$ & $1.6\!\times\!10^{-16}$ & $85.7$ & $2.1\!\times\!10^{-24}$ \\
\hline[1pt]
\end{talltblr}
\end{table}

\noindent\textbf{Player-Level Detection Performance.}
As discussed in \cref{s:player_detection}, we aggregate trajectory-level predictions and flag a player once they exceed the threshold $T$.
\cref{t:player-level_detection} summarizes player-level cheating detection for $T\in[1,5]$ for the compact case $n{=}T$, given trajectory-level rates ($\mathrm{TPR}_T = 98.98\%$, $\mathrm{FPR}_T = 2.64\%$).
Increasing $T$ sharply lowers the player-level FPR (from $6.97\times10^{-4}$ at $T{=}2$ to $1.28\times10^{-8}$ at $T{=}5$ for a single match), and aggregating two matches reduces it by further orders of magnitude, at a modest cost in recall.
Because a real match typically yields more than $T$ qualifying interactions, the operational FPR generally exceeds this compact case; we ablate the rates over the realistic range $n\in\{5,10,15,20\}$ in \cref{a:nablation}, where recall instead saturates near $100\%$.
At the observed benign mean $n{=}15$, for instance, a two-match decision at $T{=}3$ still yields $\mathrm{FPR}_P=4.4\times10^{-5}$ while retaining $\approx\!100\%$ recall, which we take as \toolname's operating point.
Developers can tune $T$ and the number of matches to balance precision and recall.

\begin{figure}
\centering
\begin{minipage}[t]{0.255\textwidth}
    \centering
    \includegraphics[width=\textwidth]{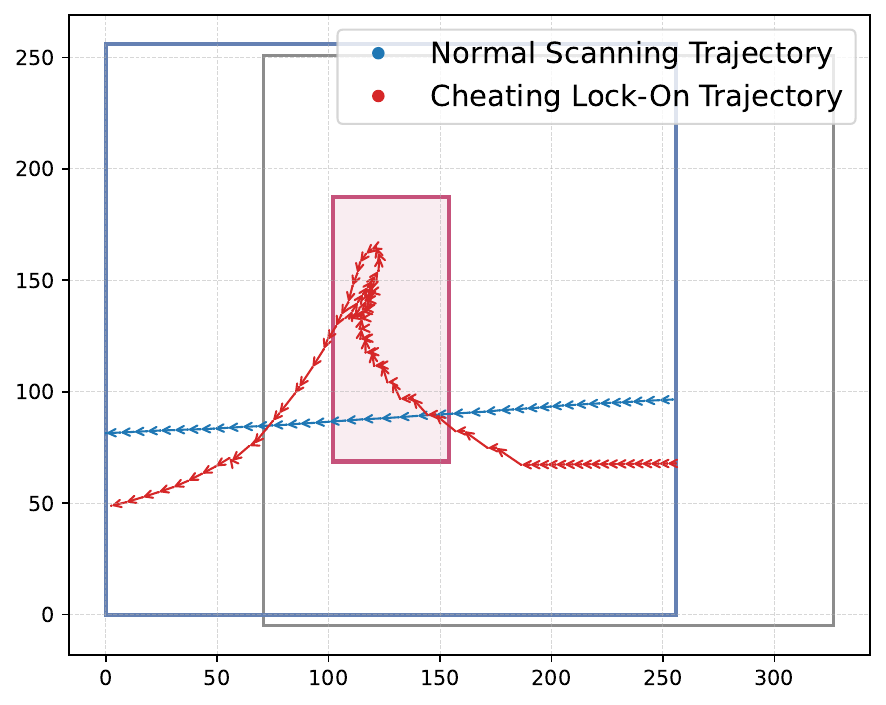}
    \subcaption{\footnotesize{$H_1$ Interaction Trajectories}}
    \label{subfig:h_1}
\end{minipage}
\begin{minipage}[t]{0.21\textwidth}
    \centering
    \includegraphics[width=\textwidth]{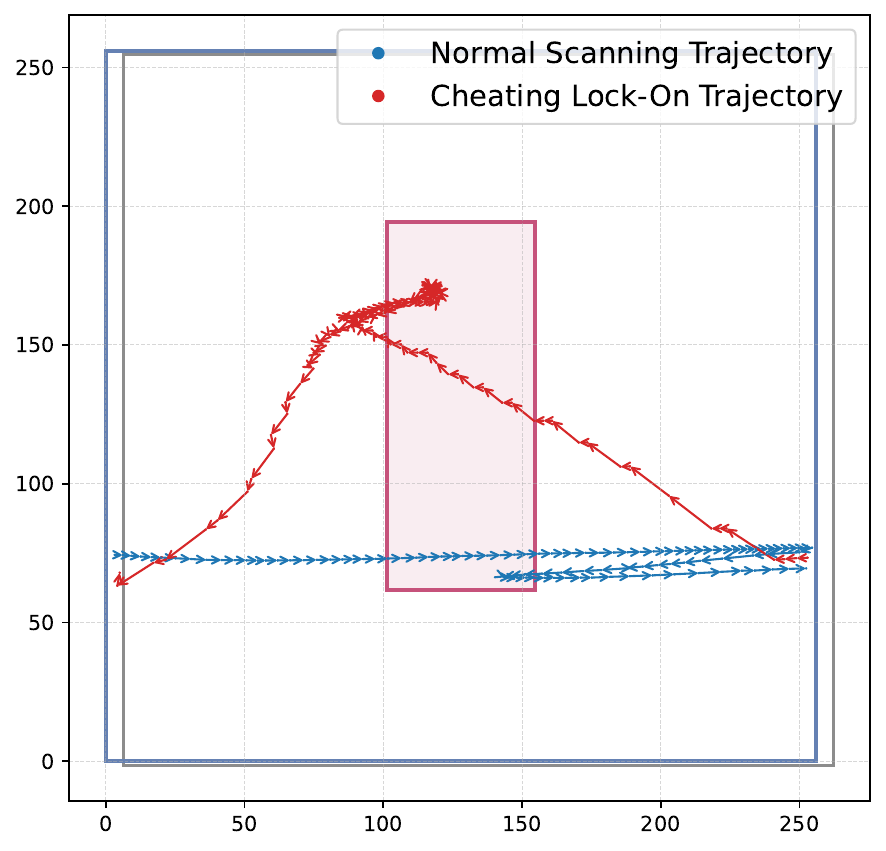}
    \subcaption{\footnotesize{$H_2$ Interaction Trajectories}}
    \label{subfig:h_2}
\end{minipage}
\caption{Typical mouse trajectories of aimbots and normal players on two distinct honeypots.}
\label{f:aiming_trajectory}
\vspace{-0.2cm}
\end{figure}

\noindent\textbf{Trajectory Visualization and Case Studies.}
\cref{f:aiming_trajectory} visualizes representative honeypot-interaction mouse trajectories for aimbot cheaters and normal players.
Red rectangles mark two honeypots ($H_1,\,H_2$); gray denotes the wall; blue outlines the Aiming Determination Area (\cref{s:detection}).
Colored dots plot the aiming trajectory with red denoting aimbot and blue denoting human trajectories.
The arrow represents the mouse moving direction.
These figures reveal a clear signal: \emph{aimbot trajectories exhibit deliberate lock\mbox{-}ons} that converge tightly on the honeypots, whereas human trajectories pass smoothly through the scene without sustained fixation.
More visualization cases can be found in \cref{a:case}.

\begin{table*}[t]
\centering
\begin{talltblr}[
    caption = {3D rendering robustness of Adversarial Camouflage Textures},
    label = {t:robustness_act},
    remark{Note} = {\emph{$\mathrm{CD}_0$} denotes camera distances $\in[0.8,\;1.0]$, while \emph{$\mathrm{CD}_1$} $\in[1.0, 1.2]$. \emph{$\mathrm{Conf}$} denotes the change in the aimbot's confidence score before and after the application of ACT. \emph{ESR} denotes Evasion Success Rate. \emph{Camera Pitch} is set to $[-2^\circ,\;2^\circ]$.}
] {
    stretch = 0.5,
    colspec = {c|c|c|c|c|c|c|c|c|c},
    rows = {font=\footnotesize},
    row{1} = {font=\tiny},
}
    \hline[1pt]
    \SetCell[c=2]{c}{\footnotesize $\mathbf{Yaw}$} &
    & $[-60^\circ,\,-45^\circ]$
    & $[-45^\circ,\,-30^\circ]$
    & $[-30^\circ,\,-15^\circ]$
    & $[-15^\circ,\,0^\circ]$
    & $[0^\circ,\,15^\circ]$
    & $[15^\circ,\,30^\circ]$
    & $[30^\circ,\,45^\circ]$
    & $[45^\circ,\,60^\circ]$ \\
    \SetCell[c=2]{c} & &
    \adjincludegraphics[width=1.4cm,trim={{0.15\width} {0.10\height} {0.15\width} {0.10\height}},clip]{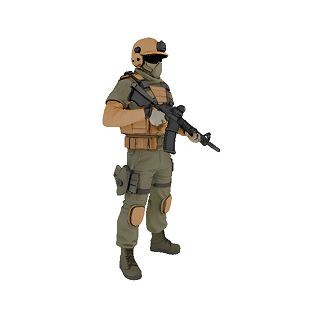} &
    \adjincludegraphics[width=1.4cm,trim={{0.15\width} {0.10\height} {0.15\width} {0.10\height}},clip]{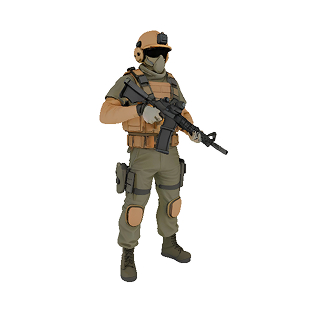} &
    \adjincludegraphics[width=1.4cm,trim={{0.15\width} {0.10\height} {0.15\width} {0.10\height}},clip]{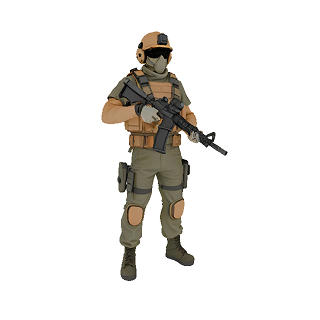} &
    \adjincludegraphics[width=1.4cm,trim={{0.15\width} {0.10\height} {0.15\width} {0.10\height}},clip]{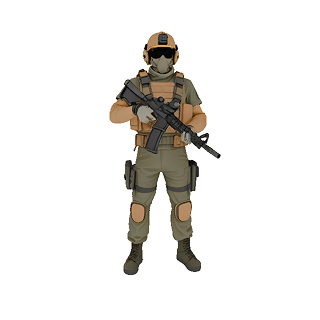} &
    \adjincludegraphics[width=1.4cm,trim={{0.15\width} {0.10\height} {0.15\width} {0.10\height}},clip]{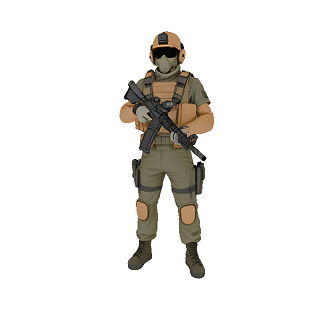} &
    \adjincludegraphics[width=1.4cm,trim={{0.15\width} {0.10\height} {0.15\width} {0.10\height}},clip]{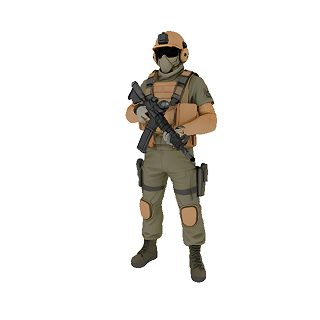} &
    \adjincludegraphics[width=1.4cm,trim={{0.15\width} {0.10\height} {0.15\width} {0.10\height}},clip]{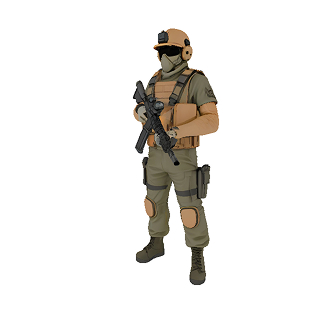} &
    \adjincludegraphics[width=1.4cm,trim={{0.15\width} {0.10\height} {0.15\width} {0.10\height}},clip]{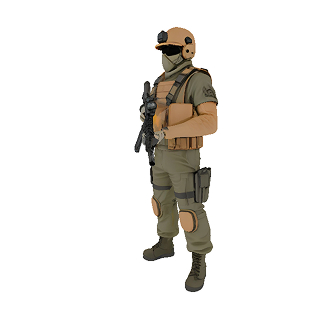} \\
    \hline
    \SetCell[r=2]{c} \textbf{$\mathbf{CD}_0$} 
    & \textbf{$\mathbf{Conf}$} & $36\% \to \textbf{3\%}$ & $45\% \to \textbf{8\%}$ & $49\% \to \textbf{8\%}$ & $61\% \to \textbf{12\%}$ & $64\% \to \textbf{13\%}$ & $58\% \to \textbf{9\%}$ & $62\% \to \textbf{15\%}$ & $68\% \to \textbf{31\%}$ \\
    & $\mathbf{ESR}$ & \textbf{99.2\%} & \textbf{94.4\%} & \textbf{95.4\%} & \textbf{94.7\%} & \textbf{91.7\%} & \textbf{92.1\%} & \textbf{88.2\%} & \textbf{66.7\%} \\
    \hline
    \SetCell[r=2]{c} \textbf{$\mathbf{CD}_1$}
    & \textbf{$\mathbf{Conf}$} & $62\% \to \textbf{17\%}$ & $60\% \to \textbf{13\%}$ & $63\% \to \textbf{17\%}$ & $71\% \to \textbf{27\%}$ & $72\% \to \textbf{22\%}$ & $70\% \to \textbf{18\%}$ & $75\% \to \textbf{26\%}$ & $76\% \to \textbf{43\%}$ \\
    & $\mathbf{ESR}$ & \textbf{81.2\%} & \textbf{83.1\%} & \textbf{82.3\%} & \textbf{70.9\%} & \textbf{78.2\%} & \textbf{82.5\%} & \textbf{72.4\%} & \textbf{39.1\%} \\
    \hline[1pt]
\end{talltblr}
\end{table*}

\begin{table}[t]
\centering
\begin{talltblr}[
    caption = {Comparison of visual-cheat detection in real-game settings. Green indicates correct predictions.},
    label = {t:real-world_cheating_detection},
    remark{Note} = {$XG$ is the state-of-the-art detector \textsc{XGuardian}~\cite{xguardian}; \textit{FPS Exp} represents total gameplay experience (hours); $I_A, I_C$ denote total honeypot interactions and cheating-specific interactions per match, respectively; $Pred$ is model prediction output, where $Pred \in \{T, F, \perp\}$. $T$ denotes a cheating classification, $F$ denotes a non-cheating classification, and $\perp$ indicates a processing failure.}
] {
    stretch = 0.5,
    colspec = {c|c|ccc|c|ccc|c},
    rows = {font=\footnotesize},
    row{3} = {font=\scriptsize}
}
\hline[1pt]
\SetCell[c=2]{c} \textbf{Player} & & 
\SetCell[c=4]{c} \textbf{No Cheating} & & & &
\SetCell[c=4]{c} \textbf{Cheating} \\
\hline
\SetCell[r=2]{c} ID & 
\SetCell[r=2]{c} FPS Exp. & 
\SetCell[c=3]{c} \toolname & & & 
XG &
\SetCell[c=3]{c} \toolname & & & 
XG \\
\hline
& & $I_{A}$ & 
$I_{C}$ & 
Pred & 
Pred &
$I_{A}$ & 
$I_{C}$ &
Pred & 
Pred \\
\hline
\textit{A} & $300\text{--}1000$ & 20 & 0 & \SetCell{bg=ForestGreen!35} F & \SetCell{bg=ForestGreen!35} F & 25 & 12 & \SetCell{bg=ForestGreen!35} T & \SetCell{bg=Red!55} F \\
\textit{B} & $>1000$ & 17 & 0 & \SetCell{bg=ForestGreen!35} F & \SetCell{bg=ForestGreen!35} F & 11 & 5 & \SetCell{bg=ForestGreen!35} T & \SetCell{bg=Red!35} $\perp$ \\
\textit{C} & $300\text{--}1000$ & 7 & 0 & \SetCell{bg=ForestGreen!35} F & \SetCell{bg=ForestGreen!35} F & 16 & 12 & \SetCell{bg=ForestGreen!35} T & \SetCell{bg=Red!55} F \\
\textit{D} & $300\text{--}1000$ & 14 & 0 & \SetCell{bg=ForestGreen!35} F & \SetCell{bg=Red!35} $\perp$ & 16 & 9 & \SetCell{bg=ForestGreen!35} T & \SetCell{bg=Red!35} $\perp$ \\
\textit{E} & $300\text{--}1000$ & 12 & 0 & \SetCell{bg=ForestGreen!35} F & \SetCell{bg=ForestGreen!35} F & 13 & 7 & \SetCell{bg=ForestGreen!35} T & \SetCell{bg=ForestGreen!35} T \\
\textit{F} & $<100$ & 10 & 0 & \SetCell{bg=ForestGreen!35} F & \SetCell{bg=Red!55} T & 10 & 5 & \SetCell{bg=ForestGreen!35} T & \SetCell{bg=Red!55} F \\
\textit{G} & $300\text{--}1000$ & 12 & 0 & \SetCell{bg=ForestGreen!35} F & \SetCell{bg=ForestGreen!35} F & 18 & 10 & \SetCell{bg=ForestGreen!35} T & \SetCell{bg=Red!35} $\perp$ \\
\textit{H} & $300\text{--}1000$ & 12 & 0 & \SetCell{bg=ForestGreen!35} F & \SetCell{bg=Red!35} $\perp$ & 23 & 12 & \SetCell{bg=ForestGreen!35} T & \SetCell{bg=Red!35} $\perp$ \\
\textit{I} & $>1000$ & 18 & 0 & \SetCell{bg=ForestGreen!35} F & \SetCell{bg=Red!35} $\perp$ & 26 & 13 & \SetCell{bg=ForestGreen!35} T & \SetCell{bg=Red!55} F \\
\textit{J} & $100\text{--}300$ & 11 & 0 & \SetCell{bg=ForestGreen!35} F & \SetCell{bg=Red!55} T & 11 & 7 & \SetCell{bg=ForestGreen!35} T & \SetCell{bg=Red!55} F \\
\textit{K} & $<100$ & 29 & 0 & \SetCell{bg=ForestGreen!35} F & \SetCell{bg=ForestGreen!35} F & 11 & 8 & \SetCell{bg=ForestGreen!35} T & \SetCell{bg=Red!55} F \\
\textit{L} & $100\text{--}300$ & 19 & 0 & \SetCell{bg=ForestGreen!35} F & \SetCell{bg=ForestGreen!35} F & 14 & 8 & \SetCell{bg=ForestGreen!35} T & \SetCell{bg=Red!55} F \\
\textit{M} & $300\text{--}1000$ & 13 & 0 & \SetCell{bg=ForestGreen!35} F & \SetCell{bg=ForestGreen!35} F & 15 & 12 & \SetCell{bg=ForestGreen!35} T & \SetCell{bg=ForestGreen!35} T \\
\textit{N} & $>1000$ & 16 & 0 & \SetCell{bg=ForestGreen!35} F & \SetCell{bg=ForestGreen!35} F & 23 & 19 & \SetCell{bg=ForestGreen!35} T & \SetCell{bg=Red!55} F \\
\textit{O} & $300\text{--}1000$ & 7 & 0 & \SetCell{bg=ForestGreen!35} F & \SetCell{bg=ForestGreen!35} F & 10 & 5 & \SetCell{bg=ForestGreen!35} T & \SetCell{bg=Red!55} F \\
\textit{P} & $<100$ & 17 & 0 & \SetCell{bg=ForestGreen!35} F & \SetCell{bg=ForestGreen!35} F & 16 & 11 & \SetCell{bg=ForestGreen!35} T & \SetCell{bg=Red!55} F \\
\textit{Q} & $100\text{--}300$ & 10 & 0 & \SetCell{bg=ForestGreen!35} F & \SetCell{bg=Red!55} T & 14 & 9 & \SetCell{bg=ForestGreen!35} T & \SetCell{bg=Red!35} $\perp$ \\
\textit{R} & $300\text{--}1000$ & 19 & 1 & \SetCell{bg=ForestGreen!35} F & \SetCell{bg=ForestGreen!35} F & 17 & 11 & \SetCell{bg=ForestGreen!35} T & \SetCell{bg=ForestGreen!35} T \\
\textit{S} & $>1000$ & 24 & 0 & \SetCell{bg=ForestGreen!35} F & \SetCell{bg=ForestGreen!35} F & 10 & 6 & \SetCell{bg=ForestGreen!35} T & \SetCell{bg=Red!55} F \\
\textit{T} & $100\text{--}300$ & 19 & 0 & \SetCell{bg=ForestGreen!35} F & \SetCell{bg=ForestGreen!35} F & 26 & 21 & \SetCell{bg=ForestGreen!35} T & \SetCell{bg=Red!55} F \\
\hline
\SetCell[c=2]{c} \textit{Average} & & 15 & 0 & & & 16 & 10 & & \\
\SetCell[c=2]{c} \textit{Overall} & & 306 & 1 & 20 & 14 & 325 & 202 & 20 & 3 \\
\hline[1pt]
\end{talltblr}
\end{table}

\subsection{Real-Game Cheating Detection Performance}
\label{e:real-world_detection}

To answer \textbf{RQ3}, we evaluate \toolname in a real-world CS2 setting and report \textsc{XGuardian}~\cite{xguardian} as a complementary reference.

\noindent\textbf{Real-world Gaming Environment Setup.}
We conducted a paid gameplay user study on our map (\cref{s:experimental_setup}) in CS2's \emph{Competitive} mode.
Twenty participants each played two matches: one without cheating and one assisted by the aimbot (configuration in \cref{a:aimbot}) to simulate cheating.
The participants were blind to the study's purpose and were recruited from a diverse pool of CS2 players with varying experience levels (from novices to veterans).
In total, we collected 40 matches (20 non-cheating, 20 cheating).
The configuration of \textsc{XGuardian} is described in \cref{a:implementation}.

\noindent\textbf{Overall Results.}
\cref{t:real-world_cheating_detection} summarizes the real-world detection statistics for \toolname.
Across all 40 matches, \toolname \textbf{correctly} classified every non-cheating and cheating match.

\noindent\emph{(1) Non-cheating matches.}
Across 20 non-cheating matches, participants triggered $I_{A}=306$ honeypot interactions (mean $=15$ per match).
Only one interaction (by player~\emph{R}) was misclassified, yielding a trajectory-level false positive rate of $1/306 \approx 3.27 \times 10^{-3}$, well below the $2.64\%$ measured on the held-out test set.
At the deployment threshold $T\ge3$, this single stray interaction never accumulates into a player-level flag: \toolname produced \emph{zero} false-flagged players across all 20 benign matches (empirical player-level FPR $0/20$).
Feeding the observed trajectory FPR into the aggregation model of \cref{s:player_detection} at the match-average $n{=}15$ projects a player-level FPR of $\approx\!1.5\times10^{-5}$ at $T{=}3$ (cf. the $n$-ablation in \cref{a:nablation}), consistent with observing no false-flagged players.
These results provide strong evidence of \toolname's precision under real-world gameplay.

\noindent\emph{(2) Cheating matches.}
Across 20 cheating matches, \toolname records $I_{A}=325$ honeypot interactions (mean $=16$), including $I_{C}=202$ cheating trajectories (mean $=10$). 
All players exceeded the threshold $T=3$ ($I_C=5$--$21$), confirming its effectiveness for real-game detection. 
The gap between $I_A$ and $I_C$ is expected, as the honeypot proximity zone (\cref{s:detection}) is recall-oriented and may include interactions where the honeypot is too distant or off-axis for aimbot lock-on.

\noindent\textbf{Complementary-reference Results.}
When applied to our honeypot dataset, \textsc{XGuardian} (\textsc{XG}) detected 3 of the 20 cheating matches and correctly classified 14 of the 20 non-cheating matches; several additional matches could not be processed successfully.
These results highlight the benefit of incorporating a visual-aimbot-specific challenge signal on honeypot maps.

As discussed in \cref{s:comparison}, however, \toolname and \textsc{XGuardian} differ substantially in scope and deployment requirements.
\textsc{XGuardian} is trained on large-scale datasets covering diverse cheat tools and configurations, including memory-based and visual aimbots as well as wallhacks.
In contrast, \toolname specifically targets visual aimbots by analyzing player trajectories relative to developer-deployed honeypots.
Because \textsc{XGuardian}'s datasets contain neither honeypot locations nor honeypot-triggered challenges, \toolname cannot be evaluated on them.
Conversely, applying the released \textsc{XGuardian} model to our controlled dataset measures its one-way compatibility under distribution shift rather than establishing a universal comparison between the two systems.
Accordingly, the results should not be interpreted as diminishing \textsc{XGuardian}'s broader applicability: unlike \toolname, it requires no deployed honeypots and is designed to detect a wider range of aim-assist and wallhack behaviors.

\subsection{Robustness}
\label{e:robustness}

We assess \toolname's robustness across various conditions.

\textit{(1) Camera Distance} ($d$): distance from camera to target (player model for ACT; honeypot for AHT). 
Default $d=1.0$; we sweep $d\in[0.8,\,1.2]$ (smaller $\Rightarrow$ closer).

\textit{(2) Camera Yaw} ($\psi$): horizontal rotation with default $\psi=0^\circ$ (target facing camera). For ACT, we sweep $\psi\in[-60^\circ,\,60^\circ]$ for wide lateral views; for AHT, $\psi\in[-20^\circ,\,20^\circ]$ suffices since honeypots lie on near-planar walls that require limited yaw range.

\textit{(3) Camera Pitch} ($\phi$): vertical rotation.
Default $\phi=0^\circ$ (level). 
For ACT we sweep $\phi\in[-2^\circ,\,2^\circ]$; for AHT we use $\phi\in[-20^\circ,\,20^\circ]$ to include steeper up/down views of wall-mounted textures.

\noindent \textbf{ACT Robustness.}
We randomly sample 1{,}280 EoR combinations per ACT from a Gaussian centered at the defaults (51{,}200 total).
\cref{t:robustness_act} reports slices over $(d,\psi)$ with fixed pitch $\phi\!\in[-2^\circ,2^\circ]$, including aimbot confidence on \emph{original} and \emph{ACT} textures as well as ESR.
ACTs are stable across most viewpoints.
At extremes (large $|\psi|$ and long distance), ESR drops due to reduced visible surface and foreshortening, suggesting room for further EoR augmentation.

\noindent \textbf{AHT Robustness.}
We sample 512 EoR combinations per AHT (122{,}880 renders total).
\cref{t:robustness_aht} summarizes honeypot confidence and DSR across representative $(d,\psi,\phi)$ ranges.
AHTs sustain uniformly high deception: all ranges exceed 95\% DSR (avg.\ 96.9\%), indicating robust decoys across various rendering conditions.

\begin{table}[t]
\centering
\begin{talltblr}[
    caption = {3D rendering robustness of AHT.},
    label = {t:robustness_aht},
    remark{Note} = {\emph{CD} denotes Camera Distance. \emph{DSR} denotes Decoy Success Rate. \emph{Confidence} denotes the aimbot's confidence score.}
] {
    stretch = 0.5,
    colspec = {c|c|c|c|c},
    rows = {font=\scriptsize},
    row{1,8} = {font=\tiny},
    rows = {valign=b}
}
    \hline[1pt]
    {\footnotesize $\mathbf{Yaw}$}
    & $[-20^\circ,\;-10^\circ]$
    & $[-10^\circ,\;0^\circ]$
    & $[0^\circ,\;10^\circ]$
    & $[10^\circ,\;20^\circ]$ \\
    
    {$\mathbf{Pitch}$ \\ {\tiny $[-20^\circ,\,0^\circ]$} \\} &
    \includegraphics[width=1.2cm]{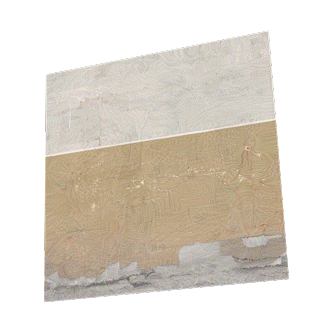} &  
    \includegraphics[width=1.2cm]{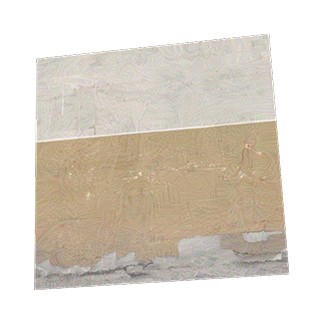} &  
    \includegraphics[width=1.2cm]{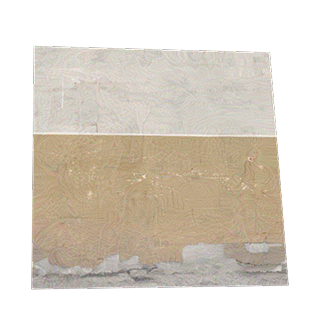} &  
    \includegraphics[width=1.2cm]{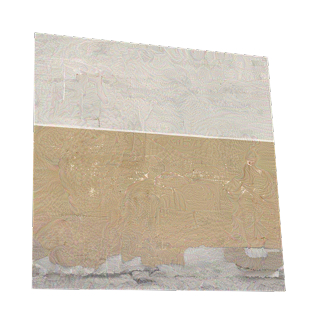} \\
    \hline
    
    $\mathbf{CD}$ &
    \SetCell[c=4]{c} $\mathbf{DSR\;/\;Confidence\;(\%)}$ \\
    \hline
    {\tiny $[0.8,\;0.9]$} & 98.4 / 79.6 & 97.2 / 79.2 & 97.2 / 79.6 & 95.4 / 76.5 \\
    {\tiny $[0.9,\;1.0]$} & 97.5 / 79.4 & 97.2 / 80.4 & 96.6 / 79.9 & 96.0 / 78.5 \\
    {\tiny $[1.0,\;1.1]$} & 98.8 / 80.2 & 97.6 / 80.6 & 96.7 / 79.6 & 96.3 / 78.0 \\
    {\tiny $[1.1,\;1.2]$} & 96.9 / 78.8 & 96.8 / 79.6 & 97.2 / 79.0 & 95.4 / 77.2 \\

    \hline[1pt]
    
    {\footnotesize $\mathbf{Yaw}$}
    & $[-20^\circ,\;-10^\circ]$
    & $[-10^\circ,\;0^\circ]$
    & $[0^\circ,\;10^\circ]$
    & $[10^\circ,\;20^\circ]$ \\
    
    {$\mathbf{Pitch}$ \\ {\tiny $[0^\circ,\,20^\circ]$} \\} &
    \includegraphics[width=1.2cm]{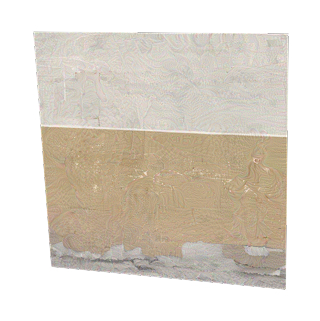} &  
    \includegraphics[width=1.2cm]{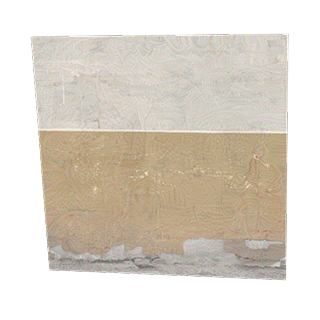} &  
    \includegraphics[width=1.2cm]{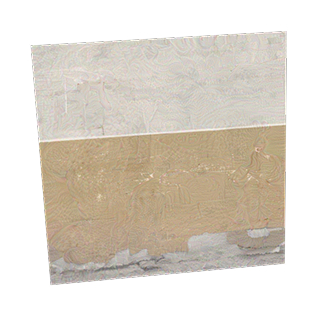} &  
    \includegraphics[width=1.2cm]{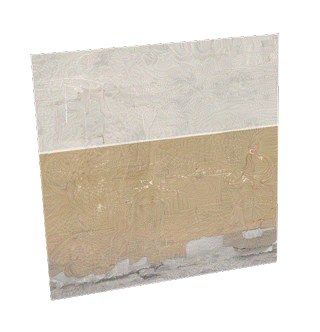} \\
    \hline
    
    $\mathbf{CD}$ &
    \SetCell[c=4]{c} $\mathbf{DSR\;/\;Confidence\;(\%)}$ \\
    \hline
    {\tiny $[0.8,\;0.9]$} & 96.8 / 79.4 & 97.4 / 79.8 & 96.1 / 79.1 & 98.1 / 80.3 \\ 
    {\tiny $[0.9,\;1.0]$} & 96.4 / 77.0 & 96.5 / 79.9 & 97.2 / 79.7 & 97.4 / 78.9 \\
    {\tiny $[1.0,\;1.1]$} & 95.7 / 76.2 & 96.9 / 79.9 & 97.1 / 80.0 & 96.9 / 78.4 \\
    {\tiny $[1.1,\;1.2]$} & 96.2 / 76.2 & 95.7 / 78.3 & 97.0 / 79.7 & 96.7 / 76.5 \\
    \hline[1pt]
\end{talltblr}
\end{table}

\subsection{Performance Overhead}
\label{e:overhead}

\begin{table}[t]
\centering
\begin{talltblr}[
    caption = {Performance overhead on adversarial texture synthesis},
    label = {t:overhead_synthesis}
] {
    stretch = 0.5,
    colspec = {Q[wd=1.1cm, c]|Q[wd=0.9cm, c]cc|cc},
    rows = {font=\footnotesize},
}
\hline[1pt]
\SetCell[r=2]{c}
& \SetCell[c=3]{c} \textbf{Time Cost} & &
& \SetCell[c=2]{c} \textbf{Storage Cost} \\
\hline
& Avg (s) & Min (s) & Max (s) 
& Resolution & Size (MB) \\
\hline
\textbf{AHT} & \textbf{12.05} & 3.08  & 42.00  & $640\times640$ & \textbf{0.83} \\
\textbf{ACT} & \textbf{74.95} & 28.12 & 119.98 & $2048\times2048$ & \textbf{7.48} \\
\hline[1pt]
\end{talltblr}
\end{table}

\noindent \textbf{\emph{Server} Offline Synthesis Overhead.}
\cref{t:overhead_synthesis} shows synthesizing one ACT averages \textbf{74.95\,s} and AHT \textbf{12.05\,s}.
Since synthesis runs offline to pre-generate texture pools (\cref{s:overview}), \toolname does not increase client overhead.
Moreover, cross-session reuse (\cref{s:texture_deployment}) amortizes this one-time cost: reusing one texture across 1{,}000 sessions reduces per-session overhead to $\mathbf{<0.1\,s}$ while maintaining sufficient variety.

\noindent \textbf{\emph{Client} Pre-Game Initialization Overhead.}
Before matches, the server transmits adversarial textures to clients (\cref{s:overview}).
Small average payloads---\textbf{0.83\,MB} per AHT and \textbf{7.48\,MB} per ACT---ensure transfer and asset installation introduce negligible loading time.

\noindent \textbf{\emph{Client} In-Game Runtime Defense Overhead.}
\toolname incurs \emph{near-zero} runtime overhead (\cref{t:overhead_comparison}).
Adversarial textures are standard assets that \emph{replace} existing ones, introducing no additional per-frame computation beyond normal rendering paths.
Although file sizes differ slightly (typically smaller than the originals), we observed no measurable latency or FPS degradation.

\noindent \textbf{\emph{Server} Post-Game Detection Overhead.}
Evaluated on 40 real-world matches (\cref{e:real-world_detection}), end-to-end post-game detection averages \textbf{8.46\,s} per player per match (\textbf{8.33\,s} for trajectory extraction, \textbf{0.13\,s} for cheating inference).
This low computational overhead makes \toolname suitable for production-level deployment.

\subsection{Real-World Gaming Impact Analysis}
\label{e:gaming_impact}

We conducted a questionnaire-based user study to assess \toolname's impact on players in real-world gaming scenarios, covering \emph{Perceptibility}, \emph{Indistinguishability}, and \emph{Naturalness}, including 113 participants with diverse FPS experience (detailed in \cref{a:user_study}).

\noindent\textbf{Perceptibility}.
We examine its impact on human perceptibility by asking whether (i) ACT makes player models imperceptible and (ii) AHT misleads them into noticing a human-like target on a wall.
Overall, participants reported \emph{no loss of perceptibility} and \emph{no misleading salience}.

\begin{figure}[h]
\centering
\begin{tabular}{>{\raggedright\arraybackslash}m{0.08\linewidth} >{\centering\arraybackslash}m{0.82\linewidth}}
\textbf{\small AHT}
& \includegraphics[width=\linewidth]{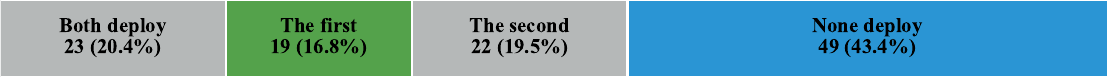} \\
\textbf{\small ACT}
& \includegraphics[width=\linewidth]{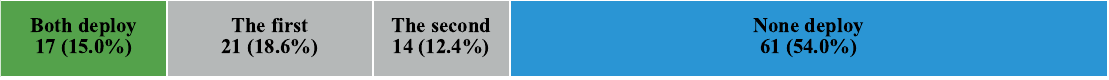} \\
\end{tabular}
\caption{Results of the user study on indistinguishability}
\label{f:indistinguish}
\end{figure}

\noindent\textbf{Indistinguishability}.
We tested whether participants could distinguish adversarial textures from the originals.
After a brief warm-up with a side-by-side comparison image, participants answered two multiple-choice questions for ACT and AHT with four options:
(i) all targets deploy \toolname, (ii) only the first, (iii) only the second, and (iv) none deploy.
The correct answers for the two items were (ii) and (i) (green in \cref{f:indistinguish}), respectively.
The results show that most participants selected ``none deploy,'' indicating that the adversarial textures were perceived as visually consistent with the originals.
The remaining responses, including the correct options, were dispersed in a near-uniform manner, suggesting that \toolname is difficult to distinguish from standard textures.

\begin{figure}[h]
\centering
\includegraphics[width=\linewidth]{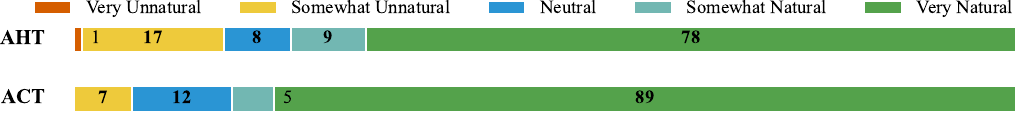} \\
\caption{Results of the user study on naturalness}
\label{f:natural}
\end{figure}

\noindent\textbf{Naturalness}.
Finally, we evaluated naturalness by presenting paired targets: one with \toolname and one without.
The participants were asked to compare the \toolname-applied model (both ACT and AHT) to the original.
As illustrated in \cref{f:natural}, most participants judged the two as equally natural.

\subsection{Transferability}
\label{e:transferability}

\begin{table}[t]
\centering
\begin{talltblr}[
    caption = {Cross-model transferability performance of AHT (\%).},
    label = {t:transferability},
    remark{Note} = {Parentheses show the change from baseline; gains are green and regressions are red. Baseline uses YOLOv5n as proxy, ENS+MI+TI uses YOLOv5n/s.}
]{
    width = \linewidth,
    stretch = 0.5,
    colspec = {X[1.4,l]|X[0.8,c]X[0.8,c]|X[c]X[c]},
    row{1,2} = {font=\footnotesize\bfseries},
    row{3-Z} = {font=\footnotesize},
    hline{1,Z} = {1pt},
    hline{3} = {0.5pt},
}
    \SetCell[r=2]{c} Aimbot Model & \SetCell[c=2]{c} \toolname (baseline) & & \SetCell[c=2]{c} \toolname (ENS+MI+TI) & \\
& DSR & Conf. & DSR & Conf. \\
YOLOv5n & 98.4 & 93.6 & 100.0 \textcolor{NiceGreen}{(+1.6)} & 95.2 \textcolor{NiceGreen}{(+1.6)} \\
YOLOv5s & 93.5 & 79.8 & 100.0 \textcolor{NiceGreen}{(+6.5)} & 95.0 \textcolor{NiceGreen}{(+15.1)} \\
YOLOv5m & 82.6 & 68.1 & 91.9 \textcolor{NiceGreen}{(+9.3)} & 82.0 \textcolor{NiceGreen}{(+13.9)} \\
YOLOv5l & 81.7 & 65.0 & 85.3 \textcolor{NiceGreen}{(+3.6)} & 72.2 \textcolor{NiceGreen}{(+7.1)} \\
YOLOv8n & 92.2 & 76.4 & 99.2 \textcolor{NiceGreen}{(+7.0)} & 82.8 \textcolor{NiceGreen}{(+6.3)} \\
Faster R-CNN & 84.3 & 74.6 & 93.3 \textcolor{NiceGreen}{(+9.0)} & 90.3 \textcolor{NiceGreen}{(+15.7)} \\
NanoDet & 19.9 & 10.5 & 38.9 \textcolor{NiceGreen}{(+19.0)} & 21.6 \textcolor{NiceGreen}{(+11.1)} \\
\hline[dashed]
\textbf{Average} & 78.9 & 66.9 & \textbf{86.9} \textcolor{NiceGreen}{(+8.0)} & \textbf{77.0} \textcolor{NiceGreen}{(+10.1)} \\
\end{talltblr}
\end{table}

We evaluate cross-model transferability using \emph{adversarial honeypot textures (AHTs)}, since reliable decoying is critical for detecting aimbots in the wild.
It is widely recognized that high transferability of adversarial attacks remains difficult~\cite{CVPRW19-fooling,cai2022context,wei2018transferable}.

We evaluate \toolname's transferability under two optimization settings: a baseline setting that uses YOLOv5n as the proxy model, and an enhanced setting that adopts an ensemble proxy of YOLOv5n and YOLOv5s combined with momentum iterative and translation-invariant optimization (ENS+MI+TI, \cref{s:aht}).
For each setup, we synthesize 12 AHTs over 120 fixed optimization iterations without early stopping and evaluate each AHT on 256 EoR samples.
The resulting baseline DSR of 93.5\% differs from the 96.9\% reported in \cref{e:defense_effectiveness} because the latter are early-stopped once the proxy is fooled; both of them use YOLOv5n as the proxy model and YOLOv5s as the aimbot model.
As \cref{t:transferability} shows, ENS+MI+TI improves average DSR and confidence by 8.0 and 10.1 percentage points, respectively.
Every model improves on both metrics under ENS+MI+TI, with the two proxy-family models (YOLOv5n/s) reaching 100.0\% DSR.
Large DSR gains also appear on Faster R-CNN (+9.0 points) and NanoDet (+19.0 points), the two non-YOLOv5 architectures, although NanoDet remains the hardest target.

\section{Discussion}
\label{s:discussion}

\subsection{Adaptive Attack}
\label{s:adaptive_attack}

As with any heuristic- or learning-based defense, \toolname may be subject to adaptive attacks when its design is known.
Nevertheless, its complementary mechanisms constrain straightforward evasion strategies and increase the attacker's cost in realistic settings.

\noindent\textbf{Threshold adjustment provides no simple escape.}
ACT reduces the aimbot's detection confidence for real players from 62.3\% to 16.0\%, whereas AHT raises its confidence for honeypots to 79.3\%.
These distributions cannot be separated by a single threshold: lowering the threshold to recover ACT-protected players also increases honeypot lock-on, while raising it to avoid honeypots further suppresses detection of real players.

\noindent\textbf{Retraining is expensive and perishable.}
First, EoR-based optimization makes adaptation from a single frame ineffective, requiring attackers to collect and label each texture across diverse viewpoints and rendering conditions.
Second, in-memory texture delivery limits offline extraction; collecting sufficient observations from live gameplay may itself induce honeypot interactions that are recorded in trajectory logs.
Third, randomized initialization seeds, per-session texture selection, and periodic rotation reduce the useful lifetime of harvested training data and require attackers to repeat the adaptation process.

\subsection{State-of-the-art Comparison}
\label{s:comparison}

\noindent\textbf{Comparison with the post-hoc behavior detector \textsc{XGuardian}.}
\textsc{XGuardian}~\cite{xguardian} trains its detection model on large-scale, real-world replay data.
Both \textsc{XGuardian} and \toolname analyze player's pitch and yaw data extracted from replay logs.
However, \toolname additionally requires maps on which visual honeypots can be deployed.
\textsc{XGuardian} provides broad coverage of memory-based and visual aimbots as well as wallhacks.
In contrast, \toolname specifically targets visual aimbots through a visual challenge-response mechanism.
The two systems are therefore complementary rather than interchangeable: \textsc{XGuardian} offers broader cheat coverage, whereas \toolname provides a targeted detection signal for visual aimbots in honeypot-equipped environments.

\noindent\textbf{Comparison with the proactive runtime defense Invisibility Cloak.}
Both systems aim to prevent visual aimbots from detecting legitimate players, but use fundamentally different approaches.
Invisibility Cloak~\cite{cloak} applies perturbations directly to display-space frames at runtime.
By contrast, \toolname's ACT consists of 3D-robust textures that are generated offline, attached statically to player models, and rendered as ordinary game assets, thereby introducing no runtime computational overhead (\cref{t:overhead_comparison}).
The two approaches also impose different deployment costs.
Invisibility Cloak requires modifications to the client-side rendering pipeline and incurs additional runtime computation.
\toolname requires honeypot-deployed maps to support its visual challenge-response detection mechanism, though the deployment effort is minimal: adversarial textures are standard assets that simply replace their original counterparts, leaving the rendering pipeline untouched.

\section{Conclusion}

This paper presents \toolname, an anti-visual-aimbot framework that can not only proactively interfere with aimbot behavior at runtime without introducing performance overhead, but also provide honeypot-based detection signals for explainable cheating detection.
Our real-game evaluations demonstrate a practical, robust, and deployable anti-visual-cheat pipeline that counters vision-based aimbots without game-engine modifications or runtime cost and offers effective and efficient offline cheating detection.

\begin{acks}
We thank our shepherd and anonymous reviewers for their insightful comments.
We thank Jun Yu and Shaofei Li for their help on our experiments.
This paper was edited for grammar using ChatGPT and Gemini.
This work was supported by Key R\&D Program of Shandong Province, China(No. 2026CXPT269, 2024CXGC010114, 2025CXPT085), National Natural Science Foundation of China under Grant(No. 62372268), Shandong Provincial Natural Science Foundation, China (No. ZR2022LZH013, No. ZR2021LZH007).
\end{acks}

\bibliographystyle{ACM-Reference-Format}
\bibliography{paper}

\appendix

\section{Ethics considerations}

The Institutional Review Board (IRB) approved this human-subjects study, and all participants gave informed consent.
We minimized risk, ensured equitable treatment, and collected only self-reported FPS gameplay hours to assess robustness across experience levels; no sensitive data were recorded.
All experiments involved only enrolled participants on a private game server, preventing bystander exposure.
We neither developed nor endorse the evaluated aimbot; it was used only under controlled conditions for research purposes.

\section{Open Science}

We provide an artifact package for \toolname at \cite{artifact}.
However, some additional artifacts cannot be fully released due to legal, ethical, and safety constraints.

\textit{(1) Raw CS2 game assets.}
Our experiments use official CS2 game textures. These assets are copyrighted by the game vendor and cannot be redistributed. Instead, we provide asset names, material categories, where appropriate, synthetic replacement textures, and scripts/configurations that allow reviewers with a licensed local installation to reproduce the asset-processing workflow.

\textit{(2) Raw user-study replay logs.}
The real-game study involved human participants. Although we collected no personally identifying information from participants directly, the game engine writes account-level identifiers (e.g., Steam IDs) into replay logs, alongside fine-grained gameplay behavior and timing traces. To protect participant privacy and comply with our ethics protocol, we therefore do not release the replay logs in their raw form. Instead, we will release a de-identified version of the logs after manually scrubbing the embedded identifiers, together with the derived honeypot-interaction trajectories, labels, aggregate statistics, and scripts sufficient to reproduce the reported detection metrics.

\section{Implementation}
\label{a:implementation}

\toolname comprises 3{,}864 lines of Python and 445 lines of Go.
The core differentiable rendering pipeline is built on PyTorch3D~\cite{pytorch3d}.
For replay analysis, we use demoinfocs~\cite{demoinfocs} to extract player actions at tick granularity in Counter-Strike~2 (CS2). All experiments were conducted on a Windows 11 PC equipped with an Intel Core i7-13700K CPU, 64~GB RAM, and an NVIDIA GeForce RTX~5090 GPU.

\noindent \textbf{Map Construction and Honeypot Deployment.}
We construct a custom CS2 map using Valve's Hammer editor~\cite{hammer}, with a layout inspired by the classic CS1.6 map \texttt{iceworld}.
The map contains 16 deployed AHTs placed at diverse scene locations to evaluate honeypot-triggered aimbot behavior under different viewing angles, distances, and player movements.
We use a custom map rather than official CS2 maps because official map source files and internal geometry metadata, such as object coordinates and dimensions, are not publicly accessible.
This metadata is necessary for computing the 3D honeypot locations and their screen-space projections during replay-based detection.
This requirement is not a fundamental limitation of our method: the detection pipeline is map-agnostic and can be integrated into any map whose internal geometry data is available to developers.

\noindent \textbf{Baseline Patch.}
To compare \toolname against the state-of-the-art baseline, we modify XGuardian~\cite{xguardian} to support our replay-file format, since its released implementation is not directly compatible with our data pipeline. The modification is minimal: XGuardian primarily uses pitch and yaw traces for detection, and these signals are fully preserved in our replay files. We therefore reuse XGuardian's released detection model and evaluate it on our dataset.

\section{Replay Logs of CS2}
\label{a:replay}
Replay logs of CS2 record each player's state at 64~ticks/s, including positions, view angles, weapon state, health, armor, and related metadata.
Typical logs are roughly 100--300\,MB and can be parsed within seconds (see our performance overhead evaluation~\cref{e:overhead}), ensuring efficient offline analysis.
Most FPS games (e.g., CS2, Overwatch) natively provide post-match replays, so our method introduces no additional storage overhead.
Moreover, our approach operates directly on standard replay logs and requires no extra instrumentation.
Among the recorded fields, we primarily use time (tick), position $(\mathrm{X}, \mathrm{Y}, \mathrm{Z})$, view angles (\textit{Yaw}, \textit{Pitch}), and alive status.

\section{Scalability to 3D Asset Generation Pipeline}
\label{a:tripo}

To assess \toolname's scalability to advanced 3D asset generation pipelines~\cite{tripo,meshy,hyer3d,aistudio}, we choose Tripo~\cite{tripo} (version v2.5) to synthesize adversarial camouflage textures.
Specifically, we generate four player models with four distinct textures, including two male and two female characters.
A representative prompt is shown below.

\begin{promptbox}
"Low-poly full-body soldier for an FPS game ($2-3$ k tris), stylized but readable geometry, flat-shaded color blocks, lightweight tactical vest, helmet, gloves and boots. Neutral combat stance: both hands firmly gripping a modern assault rifle across the chest (rifle as a separate mesh). Single $1024 \times 1024$ texture atlas, clean UVs, origin at feet, facing +Z, ready for realtime rigging."
\end{promptbox}

\section{Case Study on Real-Game Honeypot Interaction Mouse Trajectory}
\label{a:case}

\begin{figure}
\begin{minipage}[t]{0.251\textwidth}
    \centering
    \includegraphics[width=\textwidth]{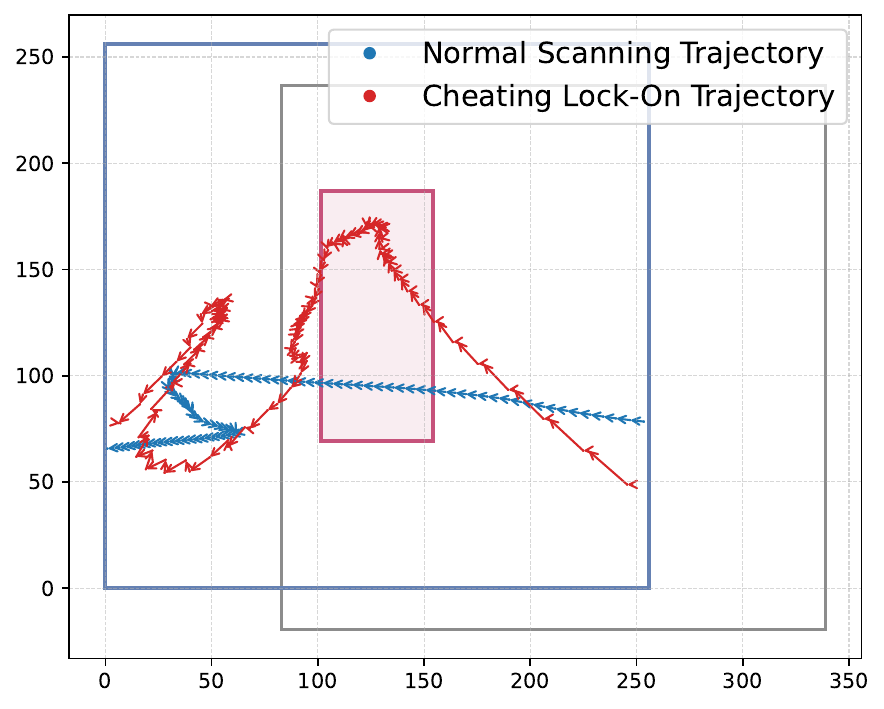}
    \subcaption{\footnotesize{$H_3$ Interaction Trajectories}}
    \label{subfig:h_3}
\end{minipage}
\begin{minipage}[t]{0.20\textwidth}
    \centering
    \includegraphics[width=\textwidth]{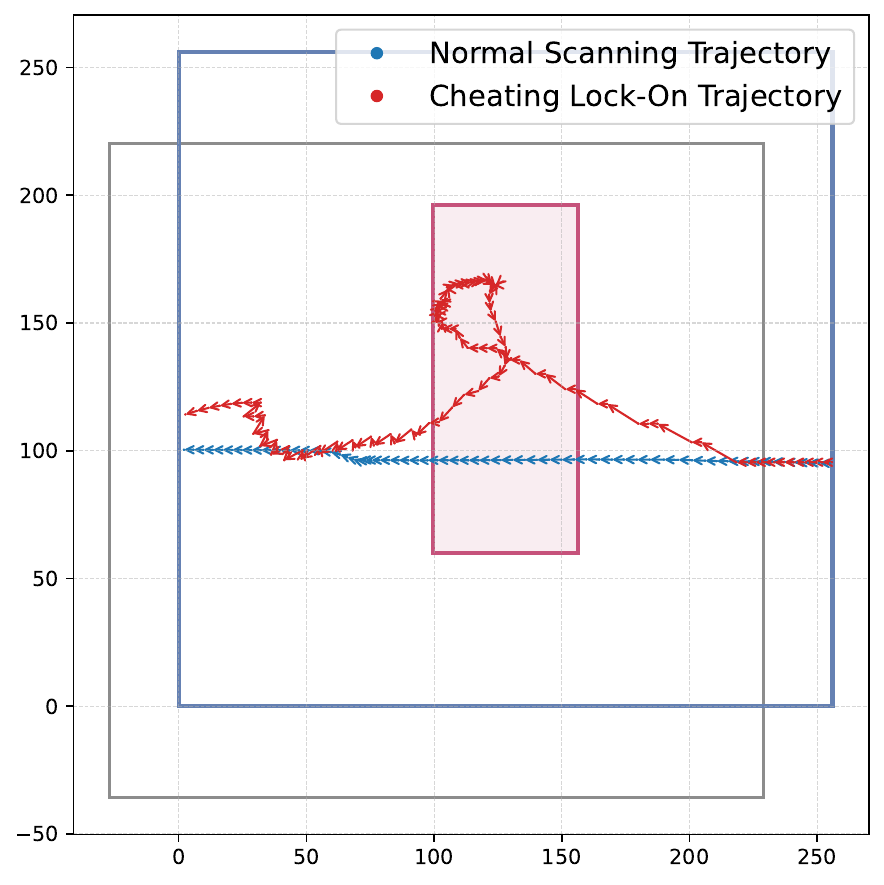}
    \subcaption{\footnotesize{$H_4$ Interaction Trajectories}}
    \label{subfig:h_4}
\end{minipage}
\caption{Typical mouse trajectories of aimbots and normal players on two distinct honeypots.}
\label{f:aiming_trajectory2}
\vspace{-0.2cm}
\end{figure}

\cref{f:aiming_trajectory2} shows two additional real-game cases of honeypot-interaction mouse trajectories from aimbot-assisted and benign players, complementing the examples in \cref{e:trajectory_identification}.

\begin{table*}[t]
\centering
\begin{talltblr}[
    caption = {Player-level detection: ablation of $\mathrm{TPR}_P$ and $\mathrm{FPR}_P$ over the number of per-match honeypot-interaction opportunities $n$ (per-trajectory $\mathrm{TPR}_T=98.98\%$, $\mathrm{FPR}_T=2.64\%$). The operational benign mean is $n{=}15$ (observed in \cref{e:real-world_detection}). },
    label = {t:fpr_ablation},
    remark{Note} = {$\mathrm{TPR}_P$ is in \%. A $k$-match decision requires a flag in all $k$ matches, i.e., the per-match probability raised to the $k$. $\mathrm{FPR}_P$ grows monotonically with $n$, whereas $\mathrm{TPR}_P$ saturates at $\approx\!100\%$ for $n\ge10$ and only drops in the degenerate $n{=}T$ corner (e.g., $n{=}T{=}5$). A $\mathrm{TPR}_P$ entry of $100$ denotes a value rounding to $100.0\%$.}
]{
    stretch = 0.5,
    colspec = {Q[wd=0.6cm, c]|Q[wd=0.9cm, c]|cccc|cccc|cccc},
    colsep = 2.5pt,
    rows = {font=\scriptsize},
    row{1,2} = {font=\scriptsize},
}
\hline[1pt]
\SetCell[r=2]{c} {\boldmath$T$} & \SetCell[r=2]{c} {\bf Metric}
& \SetCell[c=4]{c} {\bf One-Match} & & & & \SetCell[c=4]{c} {\bf Two-Match} & & & & \SetCell[c=4]{c} {\bf Three-Match} & & & \\
\hline
& & $n{=}5$ & $n{=}10$ & $\mathbf{n{=}15}$ & $n{=}20$ & $n{=}5$ & $n{=}10$ & $\mathbf{n{=}15}$ & $n{=}20$ & $n{=}5$ & $n{=}10$ & $\mathbf{n{=}15}$ & $n{=}20$ \\
\hline
\SetCell[r=2]{c} 1 & $\mathrm{TPR}_P$ & $100$ & $100$ & $\mathbf{100}$ & $100$ & $100$ & $100$ & $\mathbf{100}$ & $100$ & $100$ & $100$ & $\mathbf{100}$ & $100$ \\
 & $\mathrm{FPR}_P$ & $1.3\!\times\!10^{-1}$ & $2.3\!\times\!10^{-1}$ & $\mathbf{3.3\!\times\!10^{-1}}$ & $4.1\!\times\!10^{-1}$ & $1.6\!\times\!10^{-2}$ & $5.5\!\times\!10^{-2}$ & $\mathbf{1.1\!\times\!10^{-1}}$ & $1.7\!\times\!10^{-1}$ & $2.0\!\times\!10^{-3}$ & $1.3\!\times\!10^{-2}$ & $\mathbf{3.6\!\times\!10^{-2}}$ & $7.1\!\times\!10^{-2}$ \\
\hline
\SetCell[r=2]{c} 2 & $\mathrm{TPR}_P$ & $100$ & $100$ & $\mathbf{100}$ & $100$ & $100$ & $100$ & $\mathbf{100}$ & $100$ & $100$ & $100$ & $\mathbf{100}$ & $100$ \\
 & $\mathrm{FPR}_P$ & $6.6\!\times\!10^{-3}$ & $2.7\!\times\!10^{-2}$ & $\mathbf{5.8\!\times\!10^{-2}}$ & $9.7\!\times\!10^{-2}$ & $4.4\!\times\!10^{-5}$ & $7.4\!\times\!10^{-4}$ & $\mathbf{3.4\!\times\!10^{-3}}$ & $9.4\!\times\!10^{-3}$ & $2.9\!\times\!10^{-7}$ & $2.0\!\times\!10^{-5}$ & $\mathbf{2.0\!\times\!10^{-4}}$ & $9.1\!\times\!10^{-4}$ \\
\hline
\SetCell[r=2]{c} 3 & $\mathrm{TPR}_P$ & $100$ & $100$ & $\mathbf{100}$ & $100$ & $100$ & $100$ & $\mathbf{100}$ & $100$ & $100$ & $100$ & $\mathbf{100}$ & $100$ \\
 & $\mathrm{FPR}_P$ & $1.8\!\times\!10^{-4}$ & $1.9\!\times\!10^{-3}$ & $\mathbf{6.6\!\times\!10^{-3}}$ & $1.5\!\times\!10^{-2}$ & $3.1\!\times\!10^{-8}$ & $3.7\!\times\!10^{-6}$ & $\mathbf{4.4\!\times\!10^{-5}}$ & $2.2\!\times\!10^{-4}$ & $5.5\!\times\!10^{-12}$ & $7.1\!\times\!10^{-9}$ & $\mathbf{2.9\!\times\!10^{-7}}$ & $3.4\!\times\!10^{-6}$ \\
\hline
\SetCell[r=2]{c} 4 & $\mathrm{TPR}_P$ & $99.9$ & $100$ & $\mathbf{100}$ & $100$ & $99.8$ & $100$ & $\mathbf{100}$ & $100$ & $99.7$ & $100$ & $\mathbf{100}$ & $100$ \\
 & $\mathrm{FPR}_P$ & $2.4\!\times\!10^{-6}$ & $9.0\!\times\!10^{-5}$ & $\mathbf{5.2\!\times\!10^{-4}}$ & $1.7\!\times\!10^{-3}$ & $5.7\!\times\!10^{-12}$ & $8.1\!\times\!10^{-9}$ & $\mathbf{2.8\!\times\!10^{-7}}$ & $2.8\!\times\!10^{-6}$ & $1.3\!\times\!10^{-17}$ & $7.2\!\times\!10^{-13}$ & $\mathbf{1.4\!\times\!10^{-10}}$ & $4.7\!\times\!10^{-9}$ \\
\hline
\SetCell[r=2]{c} 5 & $\mathrm{TPR}_P$ & $95.0$ & $100$ & $\mathbf{100}$ & $100$ & $90.3$ & $100$ & $\mathbf{100}$ & $100$ & $85.7$ & $100$ & $\mathbf{100}$ & $100$ \\
 & $\mathrm{FPR}_P$ & $1.3\!\times\!10^{-8}$ & $2.9\!\times\!10^{-6}$ & $\mathbf{3.1\!\times\!10^{-5}}$ & $1.4\!\times\!10^{-4}$ & $1.6\!\times\!10^{-16}$ & $8.4\!\times\!10^{-12}$ & $\mathbf{9.5\!\times\!10^{-10}}$ & $2.0\!\times\!10^{-8}$ & $2.1\!\times\!10^{-24}$ & $2.4\!\times\!10^{-17}$ & $\mathbf{2.9\!\times\!10^{-14}}$ & $2.9\!\times\!10^{-12}$ \\
\hline[1pt]
\end{talltblr}
\end{table*}

\section{Player-Level Detection Sensitivity to Interaction Count}
\label{a:nablation}

\cref{t:player-level_detection} is reported when $n{=}T$, while real matches may yield more honeypot interactions (i.e., $n > T$).
Therefore we ablate the player-level rates over $n\in\{5,10,15,20\}$, spanning the typical range around the per-match means observed in our real-game matches ($15$ benign, $16$ cheating; \cref{e:real-world_detection}).
\cref{t:fpr_ablation} reports both $\mathrm{TPR}_P$ and $\mathrm{FPR}_P$ under this ablation for one-, two-, and three-match decisions.

\noindent\emph{Effect of $n$ on FPR.}
Since a benign player is flagged whenever at least $T$ of their $n$ interactions are misclassified, $\mathrm{FPR}_P$ grows monotonically with $n$: at $T{=}3$ it rises from $1.77\times10^{-4}$ ($n{=}5$) to $6.60\times10^{-3}$ ($n{=}15$) to $1.50\times10^{-2}$ ($n{=}20$).
The degenerate corner $n{=}T$ recovers the smallest possible value $\mathrm{FPR}_T^{\,T}$ (e.g., $1.28\times10^{-8}$ at $n{=}T{=}5$), which corresponds to requiring \emph{every} interaction to be abnormal and is therefore a lower bound rather than an operational estimate.
At the realistic $n{=}15$, a single-match decision reaches $\mathrm{FPR}_P=3.08\times10^{-5}$ only at $T{=}5$; requiring the flag in \emph{two} matches attains $4.35\times10^{-5}$ already at $T{=}3$ ($9.52\times10^{-10}$ at $T{=}5$), and a \emph{three}-match decision drives it further to $2.9\times10^{-7}$ at $T{=}3$ ($2.9\times10^{-14}$ at $T{=}5$).

\noindent\emph{Effect of $n$ on recall.}
Recall moves in the opposite direction and saturates quickly: because a cheater triggers many lock-ons per match, $\mathrm{TPR}_P>99.99\%$ for every $(n,T)$ with $n\ge10$ (\cref{t:fpr_ablation}); the only appreciable drop is the degenerate small-$n$ corner (e.g., $\mathrm{TPR}_P=95.0\%$ at $n{=}T{=}5$).
Consequently, over the realistic operating range the threshold $T$ trades almost purely against false positives at negligible recall cost.
In real-world anti-cheating deployment, minimizing false positives is critical to preserve the system's trustworthiness and avoid mis-flagging legitimate players; developers can thus pick $T$ (and the number of aggregated matches) to meet a target false-positive budget while retaining $\approx\!100\%$ recall.

\section{Real-World Gaming-Impact User Study}
\label{a:user_study}

\begin{figure}[h]
\centering
\includegraphics[width=0.85\linewidth]{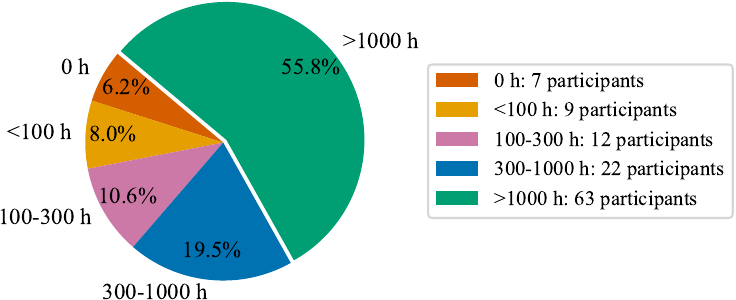} \\
\caption{FPS experience statistics of 113 participants in the questionnaire-based user study.}
\label{f:player_exp}
\end{figure}

We conducted a questionnaire-based user study to evaluate \toolname's impact on players in real-game environments.
\cref{f:player_exp} summarizes the participants' FPS experience in hours.
Most participants were experienced FPS players: 63/113 (55.75\%) reported more than 1{,}000 hours of gameplay.
At the same time, our sample also included 7/113 (6.19\%) participants with no prior FPS experience, ensuring coverage of both novice and highly active FPS players.

\section{Aimbot Configuration}
\label{a:aimbot}

We evaluate \toolname against a widely used open-source aimbot~\cite{rootkit_aimbot} ($\sim1.8\,k$ GitHub stars) that detects players with YOLOv5-series models~\cite{yolo}.
Unless stated otherwise, we use typical \emph{defaults} for the aimbot's key parameters:
(1) \textit{Aim area size:} centered $640\times 640$ pixel window.
(2) \textit{Detection confidence threshold:} default $0.4$.
(3) \textit{Aim speed (movement amplification):} default $0.4$.
(4) \textit{Headshot mode:} default \texttt{True}.
Note that the headshot mode does not affect the detection of aimbot behavior, as our approach focuses on the overall patterns of honeypot interactions rather than specific body part preferences.

\noindent\textbf{Mouse Movement Smoothing.}
The aim speed parameter is not a simple sensitivity multiplier but a \emph{mouse movement smoothing} mechanism, i.e., a proportional-controller gain applied to the cursor correction.
At each step, the aimbot computes the offset between the detected target center (adjusted by the headshot offset) and the screen center, and then issues a mouse movement of only
\begin{equation}
\Delta_{\text{mouse}} = \text{offset}_{\text{target}\rightarrow\text{crosshair}} \times \texttt{aaMovementAmp},
\end{equation}
where $\texttt{aaMovementAmp}=0.4$ by default.
Instead of snapping the crosshair onto the target in a single instantaneous jump, the aimbot moves roughly $40\%$ of the remaining offset per correction step, so the crosshair converges to the target asymptotically over multiple frames, producing a gradual, human-like approach curve rather than a discontinuous teleport.
Lower values yield smoother and slower targeting, whereas values above $1.0$ cause overshoot and jitter.
This smoothing affects only the final mouse movement command; it changes neither the detection confidence nor which target is selected.

Consequently, the aimbot we evaluate is already a \emph{trajectory-smoothing} cheat rather than an idealized instant-snap one, which is precisely the class of behavior that trajectory-based detectors struggle to flag.
\toolname is unaffected by this because its detection signal is derived from \emph{what} the cheat targets, i.e., honeypot-relative interactions, rather than \emph{how} the cursor travels there.

\end{document}